\documentclass[authoryear,review,11pt,a4paper]{elsarticle}

\usepackage{setspace}
\usepackage{fullpage}

\usepackage{color}
\usepackage{xcolor}

\usepackage{amsmath}

\usepackage{amsfonts}
\usepackage{amssymb}
\usepackage{amsthm}

\usepackage{amscd}

\usepackage{bbm}

\usepackage[ruled]{algorithm2e}

\usepackage{graphicx}

\usepackage{lscape}
\usepackage{afterpage}

\usepackage{longtable}
\usepackage{booktabs}
\usepackage{tabularx}
\usepackage{multirow}
\usepackage{float}
\usepackage{caption}
\usepackage{subcaption}

\usepackage{lineno}

\usepackage{tikz}
\usetikzlibrary{arrows.meta, positioning, fit, backgrounds}

\usepackage{hyperref}

\usepackage{listings}
\lstdefinelanguage{Renhanced}{%
  sensitive=true,
  morekeywords={function,if,else,for,while,repeat,return,break,next,in,
                library,require,source,TRUE,FALSE,NULL,NA,NaN,Inf,
                plot,print,summary},
  morestring=[b]",
  morestring=[b]',
  morecomment=[l]{\#},
  alsoletter={._},
}
\newcommand{\code}[1]{\texttt{#1}}
\newcommand{\pkg}[1]{\textbf{#1}}
\newcommand{\proglang}[1]{\textsf{#1}}

\biboptions{authoryear}

\newdefinition{rmk}{Remark}
\newproof{pf}{Proof}

\journal{Journal of Statistical Software}

\begin{document}

\begin{frontmatter}

\title{\pkg{adabay}: an \proglang{R} package for rapid evaluation and calibration of Bayesian group sequential designs across common endpoint types}

\author[TGIUK,ICTU]{Zhangyi He\fnref{fn3}}
\ead{zhe1@georgeinstitute.org.uk}


\author[BrisMath]{Feng Yu}
\ead{feng.yu@bristol.ac.uk}



\fntext[fn3]{Affiliation at time of study}

\address[TGIUK]{The George Institute for Global Health, Imperial College London, London W12 7RZ, United Kingdom}
\address[ICTU]{Imperial Clinical Trials Unit, Imperial College London, London W12 7RH, United Kingdom}
\address[BrisMath]{School of Mathematics, University of Bristol, Bristol BS8 1QU, United Kingdom}

\begin{abstract}
Bayesian group sequential designs (GSDs) combine the efficiency of frequentist GSDs with clinically interpretable probability statements and principled external evidence incorporation. However, their uptake in confirmatory trials has been held back by the cost of evaluating frequentist operating characteristics at the design stage, which often nests Markov chain Monte Carlo or another approximate posterior inference within a Monte Carlo trial-simulation loop. We introduce \pkg{adabay}, an open-source \proglang{R} package implementing a semi-simulation framework for the rapid evaluation and calibration of Bayesian GSDs. Trial data paths are simulated by Monte Carlo, while per-look posteriors and posterior tail probabilities are computed analytically or by low-dimensional deterministic quadrature. Flexible prior specification is achieved by approximating any user-specified prior with a finite mixture of conjugate components, with tail-probability diagnostics that flag inadequate approximations at the decision thresholds. The package offers a unified application programming interface for continuous, binary, count and time-to-event endpoints, supports posterior-probability decision rules with one or more efficacy and futility criteria under binding or non-binding regimes, and includes a precomputation strategy that decouples threshold calibration and look-time selection from the simulation pass. \pkg{adabay} reproduces the operating characteristics of \pkg{BATSS} and \pkg{adaptr} in the continuous and binary case studies, the analytic \pkg{gsbDesign} values in the continuous case, and the \pkg{BATSS} results in the count case, all within Monte Carlo error, while running approximately four to five orders of magnitude faster than \pkg{BATSS} and one to over two orders of magnitude faster than \pkg{adaptr} per virtual trial on eight cores. The package is distributed under an MIT licence.
\end{abstract}


\end{frontmatter}


\section{Introduction}
\label{sec:1}
Phase III confirmatory randomised controlled trials remain the gold standard for evaluating new treatments in evidence-based medicine \citep{jennison2005,pallmann2018}. They are also long, expensive, and ethically demanding, which has motivated decades of work on adaptive designs that can improve efficiency without compromising the validity of the treatment comparison. The most widely used adaptive design for confirmatory trials is the group sequential design (GSD) \citep{stevely2015,judge2021}. A GSD specifies a fixed schedule of pre-planned interim analyses on accumulating data and allows early termination when the evidence is sufficient for either efficacy or futility, while preserving the overall integrity of the comparison. Compared with fixed-sample designs, GSDs deliver substantial savings in time, cost and patient exposure. \citet{stevely2015} reported that about $68\%$ of confirmatory trials using a frequentist GSD stopped early, predominantly due to efficacy or futility.

Frequentist GSDs are well established \citep{jennison1999,wassmer2016} and routinely applied in confirmatory trials \citep[for example,][]{keh2016,combes2018,perkins2018}. Bayesian GSDs have also developed rapidly over the past two decades \citep[see][and references therein]{zhou2024}, and the regulatory environment has matured to support their adoption in confirmatory settings \citep{fda2010,fda2026}. The Bayesian formulation is motivated by clinically interpretable probability statements made directly on the parameter scale \citep{freedman1994,zhou2024} and by the ability to borrow external information through robust priors \citep[for example,][]{schmidli2014,ibrahim2015,jiang2023,fda2023,fda2026}. In a Bayesian GSD, prior beliefs about the treatment effect are encoded in a prior distribution and updated by Bayes' rule at each look, and decisions are made on quantities such as the posterior probability that the treatment is effective, the posterior probability of clinical relevance or the predictive probability of trial success \citep{spiegelhalter2004,berry2010,gsponer2014,saville2014,lee2024}.

Despite these advantages, Bayesian GSDs remain underused \citep{winkler2001,pibouleau2011,brard2017}. The principal obstacle is computational. Calibration of the design must establish that frequentist operating characteristics, in particular the type I error rate, are controlled at the levels demanded by regulators \citep{fda2010,fda2026}. Unlike spending functions, which give straightforward control of frequentist error rates in classical GSDs, Bayesian thresholds must be calibrated by Monte Carlo simulation of virtual trials. When the posterior is not analytically tractable, each simulated trial requires a posterior computation at every interim look, conventionally via Markov chain Monte Carlo (MCMC) with the associated convergence diagnostics. Nesting MCMC inside the trial-simulation loop quickly becomes prohibitive even for modest design grids \citep{lee2024}.

Existing software covers parts of this space, but no single tool provides the combination of features required for routine confirmatory use. \pkg{gsbDesign} \citep{gerber2016,gerber2024} exploits the normal--normal conjugate pair to provide closed-form per-look updates, but is restricted to normal priors and continuous endpoints. \pkg{adaptr} \citep{granholm2022,granholm2026} uses conjugate updates for Gaussian and binomial outcomes, but targets multi-arm and platform layouts and does not natively cover count or time-to-event endpoints, and its prior specification is restricted to the fixed conjugate families of its built-in models. \pkg{BATSS} \citep{couturier2024,couturier2025} replaces MCMC with the integrated nested Laplace approximation (INLA), thereby supporting a broader range of priors and of generalised-linear-model (GLM) endpoints, namely continuous, binary and count, than \pkg{gsbDesign} or \pkg{adaptr}. However, its sole model interface \code{batss.glm} is restricted to GLM likelihoods and provides no time-to-event facility, and it remains orders of magnitude slower per virtual trial than the conjugate path. Frequentist GSD packages such as \pkg{rpact} \citep{wassmer2026} and \pkg{gsDesign} \citep{anderson2026} cover a wide range of endpoints with alpha-spending machinery, but do not target Bayesian decision rules. The unmet need is a tool that simultaneously provides (i) per-look conjugate-speed posterior updates, (ii) flexibility in prior specification beyond the natural conjugate families, (iii) unified coverage of the four endpoint types most commonly encountered in confirmatory trials, and (iv) a calibration architecture that scales to large design grids.

In a companion paper \citep{he2026}, we introduced a semi-simulation framework that addresses this gap for binary endpoints. By semi-simulation, we mean that trial data paths are generated by Monte Carlo simulation as usual, but the per-look posterior of the treatment effect and its tail probabilities are obtained analytically or by low-dimensional deterministic quadrature, rather than by MCMC. Closed-form posterior updates are retained by approximating any user-specified prior that can be sampled from or evaluated numerically and adequately captured by a finite mixture of conjugate components, hereafter finite-mixture-approximable priors. The present paper extends that methodology to continuous, count, and time-to-event endpoints, and introduces \pkg{adabay}, an open-source \proglang{R} package that implements the full framework with a uniform interface across endpoint types. \pkg{adabay} makes the following contributions.

\begin{itemize}
\item A unified programming interface for Bayesian GSDs with continuous, binary, count, and time-to-event primary endpoints, together with consistent handling of absolute and relative effect scales across endpoint types.
\item Conjugacy-preserving per-look posterior updates for finite-mixture-approximable priors, with analytic or one-dimensional deterministic evaluation of the posterior tail probabilities. The mixture approximation is fitted by minimising an empirical Kullback--Leibler (KL) divergence \citep{kullback1951,dalal1983}, with tail-probability diagnostics and a runtime warning when the approximation may be unsafe for tail-driven decision rules.
\item Posterior-probability decision rules with one or more efficacy and futility criteria and binding or non-binding futility, configurable per look and per endpoint, together with the patient recruitment machinery needed to compute the expected study duration alongside the operating characteristics.
\item A precomputation strategy that decouples threshold calibration and look-time selection from the simulation pass, allowing large design grids to be evaluated at negligible additional cost after a single one-time simulation.
\item A reproducible benchmark against \pkg{gsbDesign}, \pkg{BATSS}, and \pkg{adaptr}, where available. Across the binary, continuous, and count case studies, \pkg{adabay} reproduces the relevant comparator operating characteristics within Monte Carlo error, apart from the \pkg{adaptr} type~I error rate in the continuous case (Section~\ref{sec:71}), while running approximately four to five orders of magnitude faster than \pkg{BATSS} and one to over two orders of magnitude faster than \pkg{adaptr} per virtual trial. For the time-to-event case study, no faithful comparator is constructible, since none of \pkg{gsbDesign}, \pkg{BATSS} or \pkg{adaptr} supports a time-to-event endpoint at all under the conventions used here (Table~\ref{tab:723}).
\end{itemize}

The remainder of this paper is organised as follows. Section~\ref{sec:2} gives a high-level overview of \pkg{adabay}, including the supported endpoints, decision rules and main workflow. Section~\ref{sec:3} summarises the statistical framework implemented in the package, with detailed methodological derivations deferred to the companion paper \citep{he2026}. Section~\ref{sec:4} describes the package design and user interface, focusing on the four constructor functions and the main simulation entry point. Section~\ref{sec:5} presents four case studies, one per endpoint type, with emphasis on code, output, and interpretation. Section~\ref{sec:6} states the calibration and caching architecture, contrasting direct simulation with cached grid evaluation. Section~\ref{sec:7} validates and benchmarks \pkg{adabay} against \pkg{gsbDesign}, \pkg{BATSS} and \pkg{adaptr}. Section~\ref{sec:8} discusses limitations, planned extensions and the relationship to the methodology paper \citep{he2026}.

\section{Overview of \pkg{adabay}}
\label{sec:2}
\pkg{adabay} is an open-source \proglang{R} package \citep{r2024} for the design and operating-characteristic evaluation of two-arm Bayesian GSDs. It is organised around a single workflow. The user specifies the design, including endpoint type, look schedule, and sample sizes, a prior on the per-arm outcome parameters, and a decision rule for each interim and final analysis. A single top-level function then returns Monte Carlo estimates of the operating characteristics. Most of the package, including the continuous, binary, and count trial-simulation loops and the user-facing programming interface, is written in \proglang{R}. The time-to-event (\code{"tte"}) trial simulator and the fixed-node Gauss--Legendre quadrature used for the binary and count rate-difference tails are implemented in \proglang{C++} via \pkg{Rcpp} \citep{eddelbuettel2011} for performance. The package imports \pkg{stats} from base \proglang{R} for standard quadrature and density routines, \pkg{parallel} for shared-memory parallelism, and the base \proglang{R} \pkg{graphics} package for its plotting methods. Three suggested packages are used when they are installed: \pkg{ggplot2} \citep{wickham2016} and \pkg{ggsci} for enhanced plots, and \pkg{RBesT} \citep{weber2021}, to which \code{fit\_mixture()} delegates the beta, normal and gamma mixture fits. Without \pkg{RBesT}, \code{fit\_mixture()} falls back to an internal beta-kernel EM and the normal and gamma kernels are unavailable. The source code and reproducible scripts for all case studies in Section~\ref{sec:5} are available at \url{https://github.com/trilixlab/adabay} under an MIT licence.

\paragraph{Supported endpoint types} \pkg{adabay} covers the four endpoint types most often encountered in confirmatory trials, with a closed-form per-look posterior of the treatment effect $\Delta$ for each:
\begin{itemize}
\item Continuous outcomes under a normal model, with normal--normal conjugate updates when the within-arm variance is treated as known and normal--inverse-gamma updates when it is also unknown;
\item Binary outcomes under a Bernoulli model, with beta--binomial conjugate updates per arm;
\item Count outcomes under a Poisson model on cumulative event counts, with gamma--Poisson conjugate updates per arm;
\item Time-to-event outcomes under an exponential model, with gamma--exponential conjugate updates per arm.
\end{itemize}
Treatment effects can be parameterised on the natural scale (mean, risk, and rate differences), on the relative scale (risk ratios, odds ratios, rate ratios, and hazard ratios), or on the log relative scale, with sensible defaults for each endpoint type.

\paragraph{Supported decision rules} Decisions at each look are made on posterior probabilities of the treatment effect. Each efficacy or futility rule is specified as a list of one or more criteria, all of which must be satisfied for the rule to trigger. This provides a uniform syntax for single-criterion, dual-criterion, and higher-order decision rules. The package supports both binding and non-binding futility, as well as one-sided testing in either direction.

\paragraph{Beyond conjugate priors} When the elicited prior is non-conjugate, \code{fit\_mixture()} approximates it by a finite mixture of conjugate components, fitted by minimising an empirical forward KL divergence (see Section~\ref{sec:33}). This preserves conjugacy at each look and allows posterior tail probabilities to be evaluated analytically or by one-dimensional deterministic quadrature for any finite-mixture-approximable prior. As the stopping decision relies on posterior tail probabilities, the fitted prior includes diagnostics for per-quantile tail-probability error and issues a runtime warning when the approximation may be insufficiently accurate for tail-driven decision rules.

\paragraph{Main workflow} A typical session constructs three core objects through three small constructor calls (\code{set\_design()}, \code{set\_prior()} and \code{set\_decision()}), with a fourth object created by \code{set\_accrual()} for the participant recruitment model. The recruitment model is mandatory for the time-to-event endpoint, but optional for the binary, continuous and count endpoints. These objects are then passed to the single polymorphic entry point \code{evaluate\_design()}, which has two call modes: passing an \code{adabay\_design} runs the fused simulate-and-aggregate path for a one-off single-design evaluation, while passing an \code{adabay\_cache} (built once by \code{build\_cache()} over a grid of candidate look times and effect-size thresholds) aggregates pre-computed per-look tail probabilities cheaply, so that any specific design configuration can be scored in milliseconds (see Section~\ref{sec:6}). The wrapper \code{calibrate\_design()} sweeps the cached path over a user-supplied threshold grid under target $(\alpha, 1-\beta)$ constraints and returns the design that meets the targets while minimising the expected sample size.

Both call modes of \code{evaluate\_design()} return an object of class \code{adabay\_oc} carrying the operating characteristics, while \code{calibrate\_design()} returns an \code{adabay\_calibration} object whose \code{\$best} element is an \code{adabay\_calibration\_best} object. Print methods are provided for the S3 classes \code{adabay\_design}, \code{adabay\_prior}, \code{adabay\_prior\_arm}, \code{adabay\_decision}, \code{adabay\_accrual}, \code{adabay\_cache}, \code{adabay\_calibration}, \code{adabay\_calibration\_best} and \code{adabay\_oc}; a \code{summary()} method is provided for \code{adabay\_oc}, and \code{plot()} methods for \code{adabay\_oc} and \code{adabay\_prior\_arm}. Figure~\ref{fig:21} summarises the package architecture, Table~\ref{tab:21} lists the top-level functions, Section~\ref{sec:4} documents the constructors and entry points in detail, Section~\ref{sec:47} provides a self-contained minimal working example, and Section~\ref{sec:5} presents four endpoint-specific case studies that exercise the full workflow.

\afterpage{%
\begin{landscape}
\vspace*{\fill}
\begin{figure}[!ht]
\centering
\resizebox{\linewidth}{!}{%
\begin{tikzpicture}[
  font=\small,
  >={Latex[length=2mm,width=2mm]},
  node distance = 4mm and 4mm,
  cons/.style = {rectangle, draw, rounded corners=2pt, fill=blue!8,
                 minimum width=2.5cm, minimum height=6.5mm, align=center,
                 font=\small\ttfamily},
  pre/.style  = {rectangle, draw, dashed, rounded corners=2pt, fill=blue!4,
                 minimum width=2.3cm, minimum height=6.5mm, align=center,
                 font=\small\ttfamily},
  cls/.style  = {rectangle, draw, fill=gray!10,
                 minimum width=2.2cm, minimum height=6mm, align=center,
                 font=\small\ttfamily},
  clsd/.style = {rectangle, draw, dashed, fill=gray!5,
                 minimum width=2.2cm, minimum height=6mm, align=center,
                 font=\small\ttfamily},
  ent/.style  = {rectangle, draw, rounded corners=2pt, fill=green!12,
                 minimum width=2.9cm, minimum height=6.5mm, align=center,
                 font=\small\ttfamily},
  outp/.style = {rectangle, draw, fill=orange!18,
                 minimum width=1.8cm, minimum height=6.5mm, align=center,
                 font=\small\ttfamily},
  arr/.style  = {->, semithick, draw=black!75},
  arrd/.style = {->, semithick, dashed, draw=black!60}
]

\node[cons] (sd)  {set\_design()};
\node[cons, below=of sd]  (sp)  {set\_prior()};
\node[cons, below=of sp]  (sde) {set\_decision()};
\node[cons, below=of sde] (sa)  {set\_accrual()};

\node[clsd, left=of sp]  (cpa) {adabay\_prior\_arm};
\node[pre,  left=of cpa] (fm)  {fit\_mixture()};
\draw[arr]  (fm.east)   -- (cpa.west);
\draw[arrd] (cpa.east)  -- (sp.west);

\node[cls, right=of sd]  (cd)  {adabay\_design};
\node[cls, right=of sp]  (cp)  {adabay\_prior};
\node[cls, right=of sde] (cde) {adabay\_decision};
\node[cls, right=of sa]  (ca)  {adabay\_accrual};

\draw[arr] (sd.east)  -- (cd.west);
\draw[arr] (sp.east)  -- (cp.west);
\draw[arr] (sde.east) -- (cde.west);
\draw[arr] (sa.east)  -- (ca.west);

\node[draw=black!40, dashed, rounded corners=2pt,
      fit=(cd)(cp)(cde)(ca), inner sep=2mm,
      label={[font=\scriptsize\itshape, yshift=-1mm]above:design specification}]
  (bundle) {};

\node[ent, right=7mm of bundle.east|-cd, yshift=-2mm] (st) {evaluate\_design()};
\node[ent, right=7mm of bundle.east|-sa, yshift=2mm]  (bc) {build\_cache()};

\draw[arr] (bundle.east|-st.west) -- (st.west);
\draw[arr] (bundle.east|-bc.west) -- (bc.west);

\node[outp, right=7mm of st]  (oc1)   {adabay\_oc};
\node[outp, right=7mm of bc]  (cache) {adabay\_cache};

\draw[arr] (st.east) -- (oc1.west);
\draw[arr] (bc.east) -- (cache.west);

\node[ent, right=7mm of cache, yshift=8mm]  (ed)  {evaluate\_design()};
\node[ent, right=7mm of cache, yshift=-8mm] (cal) {calibrate\_design()};

\draw[arr] (cache.east) -- ++(3mm,0) |- (ed.west);
\draw[arr] (cache.east) -- ++(3mm,0) |- (cal.west);

\node[outp, right=7mm of ed]  (oc2) {adabay\_oc};
\node[outp, right=7mm of cal] (oc3) {adabay\_calibration};

\draw[arr] (ed.east)  -- (oc2.west);
\draw[arr] (cal.east) -- (oc3.west);

\end{tikzpicture}%
}
\caption{Architecture of \pkg{adabay}. The four constructor functions \code{set\_design()}, \code{set\_prior()}, \code{set\_decision()} and \code{set\_accrual()} (left) populate the four S3 input classes (centre, grouped in the dashed \emph{design specification} box). Two top-level entry points consume the design specification directly: the fused-path call mode of \code{evaluate\_design()}, which takes an \code{adabay\_design} and returns an \code{adabay\_oc} object carrying the operating characteristics of a single design, and \code{build\_cache()}, which returns an \code{adabay\_cache} object storing the per-look posterior tail probabilities. The cache then feeds both remaining entry points: the cached-path call mode of \code{evaluate\_design()} scores a single candidate threshold combination, and \code{calibrate\_design()} searches a grid of candidate thresholds against $(\alpha,1-\beta)$ targets, scoring each grid point at marginal cost relative to the one-time simulation pass and returning an \code{adabay\_calibration} object whose \code{\$best} element carries the selected thresholds together with the underlying \code{adabay\_oc} objects. The two boxes labelled \code{evaluate\_design()} are the same function dispatched on the class of its first argument: \code{adabay\_design} (fused) versus \code{adabay\_cache} (cached). The non-conjugate-prior tributary \code{fit\_mixture()} sits on the row of \code{set\_prior()}, to its left: \code{fit\_mixture()} returns an \code{adabay\_prior\_arm} object (dashed box) which is consumed by \code{set\_prior()} through its \code{arms} argument, after which the standard constructor route to \code{adabay\_prior} resumes.}
\label{fig:21}
\end{figure}
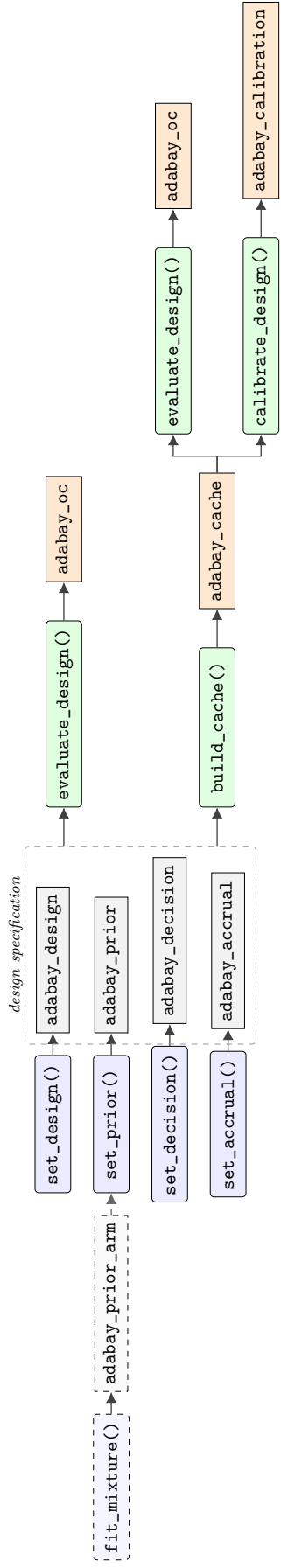
\vspace*{\fill}
\end{landscape}
}

\begin{table}[!ht]
\centering
\begin{tabularx}{\textwidth}{llX}
\toprule
Function & Returns & Purpose \\
\midrule
\code{set\_design()}        & \code{adabay\_design}     & Specify endpoint, look schedule, sample sizes and effect scale \\
\code{set\_prior()}         & \code{adabay\_prior}      & Specify per-arm conjugate or mixture-of-conjugate priors \\
\code{set\_decision()}      & \code{adabay\_decision}   & Specify per-look efficacy and futility rules \\
\code{set\_accrual()}       & \code{adabay\_accrual}    & Specify recruitment\\
\code{fit\_mixture()}       & \code{adabay\_prior\_arm} & Approximate a non-conjugate prior by a conjugate mixture \\
\code{build\_cache()}       & \code{adabay\_cache}      & Cache per-look tail probabilities over a grid \\
\code{evaluate\_design()}   & \code{adabay\_oc}         & Evaluate operating characteristics; polymorphic over \code{adabay\_design} (fused simulate-and-aggregate path) and \code{adabay\_cache} (cheap aggregation from a pre-built cache) \\
\code{calibrate\_design()}  & \code{adabay\_calibration} & Search a threshold grid under $(\alpha,1-\beta)$ targets; carries \code{\$grid} over all grid points and \code{\$best}, the selected design (\code{NULL} if no point meets both targets) \\
\code{summarise\_oc()}      & \code{data.frame}      & Tidy summary across designs in a grid \\
\bottomrule
\end{tabularx}
\caption{Top-level functions in \pkg{adabay}.}
\label{tab:21}
\end{table}

\section{Statistical framework implemented}
\label{sec:3}
This section gives the minimum statistical background needed to follow the rest of the paper. Full derivations of the per-look posterior of the treatment effect $\Delta$ for each endpoint, of the semi-simulation framework and of the conjugate-mixture prior-approximation device are given in the companion methodology paper \citep{he2026} and in the supplementary material. We use the notation of the companion paper. We consider a two-arm group sequential trial with $K$ looks indexed by $k=1,2,\ldots,K$. The arm index is $a\in\{c,t\}$, with $c$ representing the control arm and $t$ the treatment arm, and the pair of arm-specific outcome parameters is $\boldsymbol{\psi}=(\psi_c,\psi_t)$. The data accumulated up to the $k$-th look are denoted $\mathcal{D}_k$, and the posterior density of $\Delta$ at the $k$-th analysis is denoted $p_k(\delta\mid\mathcal{D}_k)$. The primary hypotheses of interest are one-sided, $H_0:\Delta\leq\Delta_{H_0}$ against $H_1:\Delta>\Delta_{H_0}$ for some prespecified value $\Delta_{H_0}$. The reverse alternative, $H_1:\Delta<\Delta_{H_0}$ is handled analogously by symmetry.

\subsection{Decision rules}
\label{sec:31}
Decisions at each analysis are based on posterior probabilities for the treatment effect \citep{lee2024,gsponer2014}. The trial stops for efficacy at the $k$-th look if all $U_k$ efficacy criteria are simultaneously met,
\begin{linenomath}
\begin{equation}
\label{eqn:311}
\mathbb{P}(\Delta>e_{k,u}\mid\mathcal{D}_k)>p_{k,u},\quad u=1,2,\ldots,U_k,
\end{equation}
\end{linenomath}
and for futility if all $V_k$ futility criteria are simultaneously met,
\begin{linenomath}
\begin{equation}
\label{eqn:312}
\mathbb{P}(\Delta<f_{k,v}\mid\mathcal{D}_k)>q_{k,v},\quad v=1,2,\ldots,V_k,
\end{equation}
\end{linenomath}
where $e_{k,u}$ and $f_{k,v}$ are user-specified effect-size thresholds and $p_{k,u}$ and $q_{k,v}$ are user-specified posterior-probability thresholds. The single-criterion case ($U_k=V_k=1$) is the most common. Multi-criterion rules \citep{gsponer2014} are also supported, for example, they can require both a high posterior probability of superiority and sufficient posterior evidence that the treatment effect exceeds a clinically relevant threshold. The trial result at the $k$-th look is summarised by
\begin{linenomath}
\begin{equation}
\label{eqn:313}
C_k=\prod_{u=1}^{U_k}\mathbbm{1}_{\{\mathbb{P}(\Delta>e_{k,u}\mid\mathcal{D}_k)>p_{k,u}\}}-\prod_{v=1}^{V_k}\mathbbm{1}_{\{\mathbb{P}(\Delta<f_{k,v}\mid\mathcal{D}_k)>q_{k,v}\}},
\end{equation}
\end{linenomath}
with $C_k=1$ for efficacy, $C_k=-1$ for futility and $C_k=0$ to continue. At the final analysis only the efficacy product is used,
\begin{linenomath}
\begin{equation}
\label{eqn:314}
C_K=2\prod_{u=1}^{U_K}\mathbbm{1}_{\{\mathbb{P}(\Delta>e_{K,u}\mid\mathcal{D}_K)>p_{K,u}\}}-1,
\end{equation}
\end{linenomath}
giving a binary trial result. Binding and non-binding futility rules are both supported. Under the usual regulatory treatment of non-binding futility \citep{fda2019}, the reported type I error rate is the rate that would be observed if the futility rule were ignored. A key engineering observation underlying \pkg{adabay} is that the ternary-valued $C_k\in\{-1,0,+1\}$ depends only on the per-look posterior tail probabilities at the threshold values $e_{k,u}$ and $f_{k,v}$. Any threshold combination $(p_{k,u},q_{k,v})$ can therefore be evaluated essentially for free once the tail probabilities have been computed, which is the property that motivates the caching architecture of Section~\ref{sec:6}.

\subsection{Per-look posterior of \texorpdfstring{$\Delta$}{Delta}}
\label{sec:32}
Under the natural conjugate prior for each endpoint, \pkg{adabay} keeps the per-look posteriors of the per-arm parameters in their conjugate parametric family. The posterior tail probability of $\Delta$ then reduces either to a standard cumulative distribution function in closed form (continuous endpoints with known variance, count and time-to-event endpoints on relative-effect scales) or to a one-dimensional deterministic computation by fixed-node Gauss--Legendre quadrature (binary endpoints, count endpoints on the difference scale) or by adaptive quadrature of a convolution integral (continuous endpoints with unknown variance). Table~\ref{tab:321} summarises the four endpoint types, the outcome model, the conjugate prior, the parameterisations of $\Delta$ supported, and the form to which the posterior tail probability reduces. The full derivations are provided in \citet{he2026} and in the supplementary material. The explicit changes of variables required for the relative-effect scales are stated in~\ref{apx:A} and~\ref{apx:B}.

\afterpage{%
\begin{landscape}
\vspace*{\fill}
\begin{table}[!ht]
\centering
\small
\begin{tabular}{p{2.2cm}p{3.8cm}p{5.6cm}p{4.0cm}p{6.2cm}}
\toprule
Endpoint & Outcome model & Conjugate prior (per arm) & Effect scale & Form of posterior tail of $\Delta$ \\
\midrule
Continuous (known $\sigma_a^2$)   & $Y_{a,i}\sim\mathcal{N}(\mu_a,\sigma_a^2)$ & $\mu_a\sim\mathcal{N}(\nu_a,\rho_a^2)$ & $\mu_t-\mu_c$ or $(\mu_t-\mu_c)/\sigma$ & Standard normal CDF \\
Continuous (unknown $\sigma_a^2$) & $Y_{a,i}\sim\mathcal{N}(\mu_a,\sigma_a^2)$ & $(\mu_a,\sigma_a^2)\sim\mathcal{NIG}(\nu_a,\kappa_a,\alpha_a,\beta_a)$ & $\mu_t-\mu_c$ or $(\mu_t-\mu_c)/\sigma$ & Convolution of two $t$ distributions \\
Binary     & $Y_{a,i}\sim\mathrm{Bernoulli}(\vartheta_a)$ & $\vartheta_a\sim\mathrm{Beta}(a_a,b_a)$ & $\vartheta_t-\vartheta_c$, $\vartheta_t/\vartheta_c$, odds ratio & 64-node 1-D Gauss--Legendre quadrature \\
Count      & $S_{a,k}\sim\mathrm{Poisson}(E_{a,k}\lambda_a)$ & $\lambda_a\sim\mathrm{Gamma}(a_a,b_a)$ & $\lambda_t-\lambda_c$, $\lambda_t/\lambda_c$, $\log(\lambda_t/\lambda_c)$ & Beta CDF (relative scales) or 1-D quadrature (difference scale) \\
Time-to-event & $T_{a,i}\sim\mathrm{Exp}(\lambda_a)$ & $\lambda_a\sim\mathrm{Gamma}(a_a,b_a)$ & $\lambda_t/\lambda_c$, $\log(\lambda_t/\lambda_c)$ & Beta CDF \\
\bottomrule
\end{tabular}
\caption{Per-endpoint conjugate model implemented by \pkg{adabay}. For the standardised mean-difference scale under the continuous endpoint, $\sigma$ denotes the common within-arm standard deviation under the assumption $\sigma_c^2=\sigma_t^2=\sigma^2$. ``Beta CDF'' refers to the standard beta cumulative distribution function applied to a deterministic transformation of $\Delta$, derived in~\ref{apx:B}. The continuous endpoint with known $\sigma_a^2$ recovers the closed-form per-look update used in \pkg{gsbDesign} \citep{gerber2016}.}
\label{tab:321}
\end{table}
\vspace*{\fill}
\end{landscape}
}

\subsection{Prior approximation by a finite conjugate mixture}
\label{sec:33}
For any finite-mixture-approximable prior $p(\psi)$, \pkg{adabay} approximates the prior by
\begin{linenomath}
\begin{equation}
\label{eqn:331}
\hat{p}(\psi)=\sum_{l=1}^{L}w_l\,h(\psi;\boldsymbol{\eta}_l),
\end{equation}
\end{linenomath}
where $h(\,\cdot\,;\boldsymbol{\eta}_l)$ is the conjugate kernel for the endpoint at hand (beta, normal, gamma or normal--inverse-gamma) and $w_l>0$ with $\sum_{l=1}^L w_l=1$. The conjugacy-preserving update of Section~\ref{sec:32} extends to this mixture without approximation. Writing $\boldsymbol{\eta}_l^{*}$ for the conjugate update of $\boldsymbol{\eta}_l$ given $\mathcal{D}_k$ and $\mathcal{L}_k(\boldsymbol{\eta}_l)$ for the marginal likelihood of $\mathcal{D}_k$ under the single-component prior $h(\,\cdot\,;\boldsymbol{\eta}_l)$, the per-look posterior is
\begin{linenomath}
\begin{equation*}
\hat{p}(\psi\mid\mathcal{D}_k)
=
\sum_{l=1}^{L}w_l^{*}\,h(\psi;\,\boldsymbol{\eta}_l^{*}),
\qquad
w_l^{*}
=
\frac{w_l\,\mathcal{L}_k(\boldsymbol{\eta}_l)}{\sum_{l'=1}^{L}w_{l'}\,\mathcal{L}_k(\boldsymbol{\eta}_{l'})}.
\end{equation*}
\end{linenomath}
That is, each component is updated by its own conjugate rule, and the mixture weights are re-weighted by the marginal-likelihood ratios \citep{dalal1983,he2026}. The posterior tail probability of $\Delta$ at any threshold is therefore the convex combination of the per-component tail probabilities, which inherit the closed-form structure of Table~\ref{tab:321}. The expectation--maximisation (EM) updates \citep{dempster1977} for the four kernels and the further extension to a product of two independent arm-wise mixtures are given in~\ref{apx:C} and in the supplementary material. The mixture parameters are fitted by minimisation of the empirical forward KL divergence
\begin{linenomath}
\begin{equation}
\label{eqn:332}
D_{KL}(p\,\|\,\hat{p})=\int p(\psi)\log\bigl(p(\psi)/\hat{p}(\psi)\bigr)\,\mathrm{d}\psi,
\end{equation}
\end{linenomath}
through an EM algorithm. The number of components $\hat{L}$ is chosen by \code{RBesT::automixfit()}, which fits $L=1,\ldots,L_{\max}$ and selects the AIC-minimising fit, when the suggested \pkg{RBesT} package is installed; when it is not, the internal beta-kernel fallback locates an elbow in the divergence decrement (default tolerance $\varepsilon=10^{-3}$) and then, if a tail tolerance is supplied, continues past that elbow one component at a time until the maximum absolute tail-probability error falls within it, returning the smallest-error fit if none does.

Forward KL is mass-covering, so a small fitted divergence indicates that the bulk of $\hat{p}$ is close to $p$, but it does not by itself guarantee accurate tail matching. This distinction matters as the stopping decision of Eq.~(\ref{eqn:313}) is driven by posterior tail probabilities, so even small percentage-point discrepancies in the relevant tails can perturb the operating characteristics. \pkg{adabay} thus reports an empirical tail-probability error table for every fitted prior and issues a runtime warning when the maximum absolute tail-probability error exceeds a tunable tolerance. These diagnostics, together with their default settings, are documented in Section~\ref{sec:42}.

\subsection{Recruitment and expected calendar duration}
\label{sec:34}
Let $n_{a,k}$ denote the cumulative number of recruited subjects in arm $a$ at the $k$-th look, and let $t_k$ denote the calendar time at which the $k$-th analysis is triggered. \pkg{adabay} supports two families of look schedule. Continuous and binary endpoints use a fixed \emph{sample-size-driven} schedule: the $k$-th look is triggered once a target cumulative pooled sample size $n_{c,k}+n_{t,k}$ has been recruited. The count endpoint uses a fixed \emph{exposure-driven} schedule: the $k$-th look is triggered once a target cumulative pooled person-time exposure has accrued, with the event count at that look a random consequence of the exposure rather than the trigger itself. For both families, when a recruitment specification is supplied via \code{set\_accrual()}, the expected calendar time at the $k$-th look follows directly from the pooled accrual rate $r$: the target (sample size or exposure) divided by $r$, plus, for the sample-size-driven schedules, the per-patient follow-up duration \code{follow\_up} supplied to \code{set\_accrual()}. No calendar time is reported when no accrual specification is supplied.

The time-to-event endpoint instead uses an \emph{event-driven} schedule: the $k$-th look is triggered once a target total number $D_k$ of events has been observed across arms, as is standard in oncology trials. Here the calendar time $t_k$ at which $D_k$ events accumulate is itself stochastic, jointly determined by the staggered Poisson recruitment process and the per-arm event hazards $\lambda_a$. A deterministic mean-field approximation to $t_k$ is available in principle, but it is systematically biased relative to the true expected hitting time of the underlying stochastic process, with the bias growing in $D_k$. \pkg{adabay} therefore does not use such an approximation to report expected duration for time-to-event designs. Instead, the expected calendar duration is obtained empirically, as the Monte Carlo mean of the realised look-time $t_{\tau^r}^r$ at each simulated trial's own stopping look $\tau^r$, over the $R$ simulated trials of Section~\ref{sec:35}. Recruitment is configured independently of the endpoint model and decision rules through \code{set\_accrual()}, as described in Section~\ref{sec:44}.

\subsection{Operating-characteristic evaluation by Monte Carlo}
\label{sec:35}
\pkg{adabay} estimates the operating characteristics of a Bayesian GSD through Monte Carlo simulation of $R$ virtual trials. For each virtual trial, the per-look posterior tail probabilities are obtained from the closed-form computations summarised in Table~\ref{tab:321} and propagated through the mixture representation of Eq.~(\ref{eqn:331}). Let $\boldsymbol{\psi}^{*}=(\psi_c^{*},\psi_t^{*})$ denote the data-generating parameter values and let $\tau^r,C^r$ denote the realised stopping time and trial result of the $r$-th virtual trial. The probabilities of stopping for efficacy and for futility at the $k$-th look are approximated by
\begin{linenomath}
\begin{equation}
\label{eqn:351}
\alpha_k^{\boldsymbol{\psi}^{*}}\approx\frac{1}{R}\sum_{r=1}^{R}\mathbbm{1}_{\{\tau^r=k,\,C^r=+1\,\mid\,\boldsymbol{\psi}^{*}\}},
\qquad
\beta_k^{\boldsymbol{\psi}^{*}}\approx\frac{1}{R}\sum_{r=1}^{R}\mathbbm{1}_{\{\tau^r=k,\,C^r=-1\,\mid\,\boldsymbol{\psi}^{*}\}}.
\end{equation}
\end{linenomath}
The overall type I error rate and power are obtained by summing the per-look stopping probabilities under data-generating values $\boldsymbol{\psi}^{H_0}$ on the boundary of $H_0$ and under the assumed alternative. The expected sample size $\mathbb{E}_{\boldsymbol{\psi}^{*}}(N)$ and expected study duration $\mathbb{E}_{\boldsymbol{\psi}^{*}}(T)$ are obtained by averaging the realised stopping times over $\{(\tau^r,C^r)\}_{r=1}^{R}$. Algorithm~\ref{alg:351} summarises the procedure, and full derivations are provided in \citet{he2026}.

\begin{algorithm}[!ht]
  \SetKwInOut{Input}{input}
  \SetKwInOut{Output}{output}
  \Input{Design with $K$ analyses and the endpoint-specific cumulative look schedule (per-arm sample sizes $\boldsymbol{n}_{a,1:K}$, $a\in\{c,t\}$, for continuous and binary; per-arm exposures $\boldsymbol{E}_{a,1:K}$ for count; pooled event counts $\boldsymbol{D}_{1:K}$ for time-to-event), decision rules of Eqs.~(\ref{eqn:311})--(\ref{eqn:314}), prior approximation $\hat{p}$, data-generating values $\boldsymbol{\psi}^{*}$, the recruitment and follow-up model of Section~\ref{sec:34}, simulation size $R$, endpoint type}
  \Output{$\boldsymbol{\alpha}_{1:K}^{\boldsymbol{\psi}^{*}}$, $\boldsymbol{\beta}_{1:K}^{\boldsymbol{\psi}^{*}}$, $\mathbb{E}_{\boldsymbol{\psi}^{*}}(N)$, $\mathbb{E}_{\boldsymbol{\psi}^{*}}(T)$}
  \Begin{
    set $\mathcal{R}=\{1,2,\ldots,R\}$ and $\tau^{r}\leftarrow K$ for $r\in\mathcal{R}$\;
    \For{$k\leftarrow 1$ \KwTo $K-1$}{
      \For{$r\in\mathcal{R}$}{
        draw the per-arm increments in $\mathcal{D}_k\setminus\mathcal{D}_{k-1}$ from the endpoint-specific data-generating distribution at $\boldsymbol{\psi}^{*}$\;
        compute the closed-form per-look posterior of $\Delta$ (Table~\ref{tab:321}), propagating the prior mixture of Eq.~(\ref{eqn:331})\;
        evaluate $C_k^r$ via Eq.~(\ref{eqn:313})\;
        if $C_k^r\neq 0$, set $\tau^r\leftarrow k$\;
      }
      update $\mathcal{R}\leftarrow\mathcal{R}\setminus\{r:C_k^r\neq 0\}$\;
    }
    \For{$r\in\mathcal{R}$}{
      draw the per-arm increments at the $K$-th look\;
      compute the closed-form per-look posterior\;
      evaluate $C_K^r$ via Eq.~(\ref{eqn:314})\;
    }
    aggregate $\boldsymbol{\alpha}_{1:K}^{\boldsymbol{\psi}^{*}}$ and $\boldsymbol{\beta}_{1:K}^{\boldsymbol{\psi}^{*}}$ from $\{(\tau^r,C^r)\}_{r=1}^{R}$ via Eq.~(\ref{eqn:351})\;
    compute $\mathbb{E}_{\boldsymbol{\psi}^{*}}(N)$ and $\mathbb{E}_{\boldsymbol{\psi}^{*}}(T)$ from $\{\tau^r\}_{r=1}^{R}$, the look schedule, and the recruitment and follow-up model of Section~\ref{sec:34}\;
  }
  \caption{The \pkg{adabay} semi-simulation procedure produces Monte Carlo estimates of the per-look stopping probabilities, expected sample size and expected study duration for a two-arm Bayesian GSD across the four supported endpoint types.}
  \label{alg:351}
\end{algorithm}

\section{Package design and user interface}
\label{sec:4}
The programming interface of \pkg{adabay} is built around four S3 constructor functions that return the four \code{adabay\_*} classes summarised in Table~\ref{tab:21}, together with the polymorphic top-level entry point \code{evaluate\_design()} and a non-conjugate prior fitter. This section introduces the constructors in the order in which they would typically be invoked: design, prior, decision, accrual, and then the fused-path call mode of \code{evaluate\_design()}. Calibration and the cached-path call mode of \code{evaluate\_design()} are deferred to Section~\ref{sec:6}.

\subsection{Design specification}
\label{sec:41}
The \code{set\_design()} function specifies the endpoint type and look schedule. Its first argument \code{endpoint} accepts one of \code{"continuous"}, \code{"binary"}, \code{"count"} or \code{"tte"}, and subsequent arguments configure the look schedule and the cumulative sample size (\code{"continuous"} and \code{"binary"} endpoints), exposure (\code{"count"} endpoint), or event counts (\code{"tte"} endpoint), pooled across both arms, at each look.
The \code{effect\_scale} argument chooses one of the supported parameterisations of $\Delta$ for the selected endpoint, with sensible defaults: mean difference for continuous, risk difference for binary, rate difference for count, and hazard ratio for time-to-event.

\begin{lstlisting}
des <- set_design(
  endpoint     = "binary",
  n_per_look   = c(760, 1520, 2280, 3040, 3800),
  effect_scale = "risk_difference",
  alternative  = "less"
)
\end{lstlisting}

The argument \code{alternative} selects the direction of the alternative hypothesis: $H_1:\Delta>\Delta_{H_0}$ if \code{"greater"}, $H_1:\Delta<\Delta_{H_0}$ if \code{"less"}, mirroring \code{stats::t.test}. The corresponding sign-flips in Eqs.~(\ref{eqn:311}) and (\ref{eqn:312}) are propagated automatically. The returned \code{adabay\_design} object carries a \code{print()} method that displays the endpoint, the look schedule, the effect scale and the direction of the alternative.

Several additional arguments configure endpoint-specific aspects of the design. \code{delta\_null} specifies the null-hypothesis effect-size value $\Delta_{H_0}$ on the chosen effect scale, defaulting to $0$ for absolute scales and $1$ for relative scales. For continuous endpoints, \code{sigma} specifies the common within-arm standard deviation of the simulated outcome data; it is required for every continuous design, since it drives the Monte Carlo data generation, and is additionally used by the normal known-variance posterior update. For count endpoints, \code{exposure\_per\_look} plays the role of \code{n\_per\_look}, giving the cumulative pooled exposure across both arms at each analysis. For time-to-event endpoints, \code{d\_per\_look} specifies the cumulative number of events at each analysis, and \code{d\_total} specifies the total target number of events. The number of looks $K$ is inferred from the length of the supplied per-look vector. \code{allocation\_ratio} is the treatment-to-control allocation ratio, which splits each pooled per-look value into per-arm targets: for continuous and binary endpoints $n_{c,k}=\mathrm{round}(\text{\code{n\_per\_look}}_k/(1+\text{\code{allocation\_ratio}}))$ and $n_{t,k}=\text{\code{n\_per\_look}}_k-n_{c,k}$, so the per-arm sizes sum exactly to the pooled value; count exposure splits the same way without rounding, and the event-count schedule for the time-to-event endpoint is already pooled. The case studies of Section~\ref{sec:5} demonstrate the use of these arguments individually.

\subsection{Prior specification}
\label{sec:42}
Conjugate priors are specified directly through \code{set\_prior()}. Non-conjugate priors are first approximated by \code{fit\_mixture()} (Section~\ref{sec:33}) and then passed to the same constructor.

\begin{lstlisting}
## Conjugate beta priors on the per-arm response rates
pri_conj <- set_prior(
  endpoint = "binary",
  arms     = list(c = list(family = "beta", a = 1, b = 1),
                  t = list(family = "beta", a = 1, b = 1))
)

## A non-conjugate logit-normal prior on the control rate, approximated
## by a beta mixture with a default KL tolerance of 1e-3 and a tail
## tolerance of 5e-3
pri_lnorm <- fit_mixture(
  endpoint = "binary",
  arm      = "c",
  prior    = function(theta) dnorm(log(theta / (1 - theta)),
                                   mean = -0.7, sd = 0.4) /
                             (theta * (1 - theta)),
  n_components_max = 5,
  tol_kl   = 1e-3,
  tol_tail = 5e-3
)
print(pri_lnorm)   # selected L, KL, per-quantile tail-error table
plot(pri_lnorm)    # empirical-sample density vs fitted mixture density
\end{lstlisting}

The \code{fit\_mixture()} routine executes the EM algorithm of Section~\ref{sec:33} for $L=1,2,\ldots,L_{\max}$, picks $\hat{L}$ by the selection rule of Section~\ref{sec:33}, and returns an \code{adabay\_prior\_arm} object that carries the fitted mixture weights, component hyperparameters and tail-error diagnostics. If the maximum absolute tail-probability error exceeds \code{tol\_tail}, a runtime warning is emitted. The \code{print()} method displays the tail-error table by quantile, while the \code{plot()} method overlays the empirical-sample density on the fitted mixture density, together with rugs at the tail thresholds.

\subsection{Decision rules}
\label{sec:43}
Decision rules are configured per look through \code{set\_decision()}. Each efficacy or futility rule is a list of one or more criteria, with all criteria required to fire jointly for the rule to trigger. Binding and non-binding futility are both supported. By default, each look adopts one efficacy criterion and one futility criterion at $\Delta_{H_0}$, but each rule can be customised per look by passing a list of lists.

\begin{lstlisting}
## Dual-criterion efficacy with non-binding futility, common across looks
## (a single-criterion or higher-order rule has the same syntactic form)
dec <- set_decision(
  efficacy = list(
    list(threshold_effect =  0.00, threshold_prob = 0.99),
    list(threshold_effect = -0.05, threshold_prob = 0.50)
  ),
  futility = list(
    list(threshold_effect =  0.00, threshold_prob = 0.90)
  ),
  futility_binding = FALSE
)
\end{lstlisting}

\subsection{Accrual model}
\label{sec:44}
The expected study duration depends on the recruitment process, which is configured independently of the endpoint model and decision rules through \code{set\_accrual()}. \pkg{adabay} accepts three recruitment models: a homogeneous Poisson process with constant pooled rate $r$ across both arms (split into per-arm rates by \code{allocation\_ratio}), a piecewise-constant Poisson process with user-supplied breakpoints (which captures site-activation profiles), and an arbitrary user-supplied callback that returns recruitment times. Only the homogeneous Poisson model currently feeds into the reported expected calendar duration and into the time-to-event endpoint's simulated recruitment timeline (Section~\ref{sec:34}); the piecewise and callback models are accepted by \code{set\_accrual()} but do not currently drive any calendar-time computation, and are not supported at all for time-to-event designs, which require a Poisson recruitment specification.

Assuming that all patients are already at risk from the start of the trial, rather than arriving over time, overestimates each patient's exposure at any given calendar time and therefore overestimates the information available at a look. \pkg{adabay} therefore simulates staggered Poisson entry explicitly rather than treating recruitment as instantaneous, and time-to-event designs accordingly require a recruitment specification (Section~\ref{sec:54}).

\begin{lstlisting}
## Homogeneous Poisson recruitment at 40 patients per month, pooled
## across both arms
acc <- set_accrual(
  model = "poisson",
  rate  = 40
)
\end{lstlisting}

For continuous, binary and count endpoints, when a recruitment specification is supplied, the expected study duration is computed analytically as the target cumulative sample size or exposure at the $k$-th look divided by the pooled accrual rate, plus the per-patient follow-up duration \code{follow\_up} of \code{set\_accrual()} for the sample-size-driven continuous and binary schedules; the exposure-driven count schedule does not add the follow-up term (Section~\ref{sec:34}). For the time-to-event endpoint, the expected study duration is obtained as the direct Monte Carlo mean, across the simulated trials, of the calendar time realised at each trial's own stopping look under the staggered Poisson recruitment timeline.
\subsection{Evaluating a single design}
\label{sec:45}
The simplest workflow runs Algorithm~\ref{alg:351} for a single design via the fused-path call mode of \code{evaluate\_design()}, with an \code{adabay\_design} as its first argument.

\begin{lstlisting}
oc <- evaluate_design(des,
  prior    = pri_conj,
  decision = dec,
  effect   = list(theta_c = 0.33, theta_t = 0.28),
  n_trials = 1e5,
  cores    = 8,
  seed     = 1L
)
print(oc)        # Per-look stopping probabilities and overall alpha/beta
plot(oc)         # Stopping-probability bar chart by look
summary(oc)$ess  # Expected sample size
\end{lstlisting}

The returned \code{adabay\_oc} object contains the per-look stopping probabilities of Eq.~(\ref{eqn:351}), the overall probabilities of crossing the efficacy and futility boundaries under the simulated effect, the expected sample size and, when a recruitment specification is supplied via \code{set\_accrual()}, the expected study duration. Under binding futility the efficacy-crossing probability is the type I error rate when the simulated effect lies on the null boundary and the power when it is the clinically relevant alternative. Under the non-binding rule used here, the efficacy-crossing probability is the realised crossing rate under actual trial conduct, whereas the reported type I error rate is the rate that would be observed if the futility rule were ignored (Section~\ref{sec:31}); the two coincide only under binding futility. Calibration over a grid of thresholds or look times is performed by the cached-path call mode of \code{evaluate\_design()} on an \code{adabay\_cache} object (Section~\ref{sec:6}), which avoids repeating the simulation pass for every candidate configuration.

\subsection{Reproducibility, testing and distribution}
\label{sec:46}
\paragraph{Parallelism and seeds} The trial-level loop in Algorithm~\ref{alg:351} is embarrassingly parallel across the $R$ virtual trials. \pkg{adabay} parallelises this loop via \code{parallel::mclapply} on POSIX platforms and \code{parallel::parLapply} on Windows, with the number of workers set by the \code{cores} argument in \code{evaluate\_design()} and \code{build\_cache()}. When \code{cores > 1}, worker-specific random-number streams are generated using L'Ecuyer's combined multiple-recursive generator \citep{lecuyer2002}. For any fixed \code{(seed, cores)} pair, the per-trial outputs are deterministic, and repeated runs with the same worker count are therefore bit-identical. Bit-identical reproducibility across different worker counts, \textit{e.g.}, matching \code{cores = 1}, \code{cores = 4} and \code{cores = 8} under the same user-level \code{seed}, is enabled by setting the \code{cross\_core\_reproducible = TRUE} argument on \code{evaluate\_design()} and \code{build\_cache()}; the simulator pass is then forced to run in a single driver process irrespective of the user-supplied \code{cores}, so the trial-level RNG draws coincide with the \code{cores = 1} stream regardless of worker count. The posterior and aggregation passes are already core-independent. The wall-clock cost of this mode is the loss of simulator parallelism, which is a small fraction of total wall-clock for binary, count and time-to-event, and a moderate fraction for continuous. The default is \code{cross\_core\_reproducible = FALSE} so that the pinned regression tests in Section~\ref{sec:75} and the worked examples in Section~\ref{sec:5} continue to run on the parallel simulator path; the \code{TRUE} mode is intended for regulatory submissions and other settings where reproducibility across hardware configurations is required.

\paragraph{Test suite} The package includes a \pkg{testthat}-based unit-test suite under \code{tests/testthat/}, exercised on every install. The suite covers each constructor (input validation and class invariants on \code{set\_design()}, \code{set\_prior()}, \code{set\_decision()} and \code{set\_accrual()}), the EM mixture fitter and its tail-probability diagnostics in \code{fit\_mixture()}, the closed-form per-look posterior of $\Delta$ against an analytic beta--binomial benchmark for the binary endpoint and against pinned reference values at frozen seeds for the remaining endpoint types of Table~\ref{tab:321}, the trial-simulation loop at small $R$ for shape and monotonicity properties, and the caching function including the bit-for-bit equivalence between cached and direct evaluation of the same threshold combination.

\paragraph{Versioning} The package follows semantic versioning, with the version of record at the time of writing tagged \code{v0.1.0} and available from \url{https://github.com/trilixlab/adabay}. It passes \code{R CMD check} under the \code{--as-cran} profile with no errors or warnings.

\paragraph{Reproducible scripts and vignettes} The four worked case studies in Section~\ref{sec:5} are reproduced by self-contained scripts in the package's \code{inst/examples/}, one for each endpoint (\code{transform2.R}, \code{adrenal.R}, \code{bnt162b2.R}, \code{checkmate141.R}). A commented standalone script, \code{replication.R}, is provided at the top level of the supplementary code folder and reproduces every numerical result in the main text, including the four case studies in Section~\ref{sec:5}, the stage-specific calibrated five-look ADRENAL design in Section~\ref{sec:63}, the cached versus direct grid-evaluation comparison in Section~\ref{sec:64}, and the comparator scaffolding for Section~\ref{sec:7}. The script is controlled by a single \code{RUN\_FULL\_PRECISION} flag, which switches between a fast smoke-test run with $R=2000$ and the manuscript simulation budgets of $R=10^{4}$ for matched comparisons and $R=10^{6}$ for high-precision runs. Calls to \pkg{BATSS}, \pkg{adaptr} and \pkg{gsbDesign} are guarded by \code{requireNamespace()} and skipped when the comparator package is absent. 

The package also includes three \pkg{knitr}-based vignettes covering the main workflows. \code{overview} sketches the package architecture and an end-to-end minimal session on the minimal ADRENAL re-design; \code{evaluation} walks through \code{evaluate\_design()} on a fixed design under both a conjugate $\mathrm{Beta}(1,1)$ prior and a non-conjugate logit-normal prior approximated by a beta mixture via \code{fit\_mixture()}, with the per-quantile tail-probability diagnostics of Section~\ref{sec:33} on the non-conjugate case. \code{calibration} walks through the design-to-calibration workflow of Section~\ref{sec:6} (one \code{build\_cache()} pass per scenario, followed by \code{calibrate\_design()} for common-across-looks thresholds and a short \code{evaluate\_design()} loop for the stage-specific Haybittle--Peto schedule in Section~\ref{sec:63}), again under both prior choices. A PDF reference manual (\code{adabay-manual.pdf}) and a static HTML reference site (top-level \code{Website/} directory of the supplementary materials) accompany the package.

\subsection{A minimal working example}
\label{sec:47}
The constructors introduced in Sections~\ref{sec:41}--\ref{sec:44}, together with the simulation entry point in Section~\ref{sec:45}, form a self-contained workflow. The example below evaluates a design with a non-conjugate prior approximated with a beta mixture, incorporates an accrual model, and returns the headline outputs of the resulting \code{adabay\_oc} object. Its cost is dominated by the simulation and threshold-sweep passes at $R=10^{5}$; Table~\ref{tab:641} reports the measured wall-clock of a comparable single \code{build\_cache()} pass at $R=10^{5}$ on eight cores. The four case-study scripts in the package's \code{inst/examples/} directory follow the same sequence for the designs of Section~\ref{sec:5}.

\begin{lstlisting}
library(adabay)

## 1. Design: 3-look binary GSD on the risk-difference scale
des <- set_design(endpoint     = "binary",
                  n_per_look   = c(1266, 2534, 3800),
                  effect_scale = "risk_difference",
                  alternative  = "less")

## 2. Prior: non-conjugate logit-normal on the control arm,
##    approximated by a beta mixture; conjugate Beta(1,1) on the treatment arm
pri_c <- fit_mixture(endpoint = "binary", arm = "c",
                     prior    = function(theta)
                       dnorm(log(theta / (1 - theta)), mean = -0.7, sd = 0.4) /
                         (theta * (1 - theta)),
                     tol_kl   = 1e-3, tol_tail = 5e-3)
print(pri_c)                        # KL, max tail error, warnings
plot(pri_c)                         # density overlay with tail rugs

pri <- set_prior(endpoint = "binary",
                 arms     = list(c = pri_c,
                                 t = list(family = "beta", a = 1, b = 1)))

## 3. Decision rule: efficacy at P(Delta < 0) > 0.99, non-binding futility.
##    futility_binding = FALSE matches the regulatory default of Section 3.1;
##    the case studies of Section 5 use futility_binding = TRUE to keep the
##    cross-package benchmarks of Section 7 on a single footing
dec <- set_decision(efficacy = list(list(threshold_effect = 0,
                                         threshold_prob   = 0.99)),
                    futility = list(list(threshold_effect = 0,
                                         threshold_prob   = 0.90)),
                    futility_binding = FALSE)

## 4. Accrual: Poisson recruitment
acc <- set_accrual(model = "poisson", rate = 100)

## 5. Simulate the uncalibrated design under H1 and inspect
oc <- evaluate_design(des, pri, dec, accrual = acc,
                      effect   = list(theta_c = 0.33, theta_t = 0.28),
                      n_trials = 1e5, cores = 8, seed = 1L)
print(oc)                           # per-look stopping probs, efficacy and futility probabilities
plot(oc)                            # stopping-probability bar chart
summary(oc)[c("alpha", "eff_prob", "fut_prob", "ess", "edur")]

## 6. Calibrate the same design against regulator targets: one cached
##    simulation pass under H0 and one under H1, then a fast threshold
##    sweep over the two caches.
cache_h0 <- build_cache(des, pri,
                        effect   = list(theta_c = 0.33, theta_t = 0.33),
                        n_trials = 1e5, cores = 8, seed = 1L,
                        threshold_grid = list(efficacy = 0, futility = 0))
cache_h1 <- build_cache(des, pri,
                        effect   = list(theta_c = 0.33, theta_t = 0.28),
                        n_trials = 1e5, cores = 8, seed = 1L,
                        threshold_grid = list(efficacy = 0, futility = 0))
cal <- calibrate_design(cache         = cache_h0,
                        cache_alt     = cache_h1,
                        alpha_target  = 0.025,
                        power_target  = 0.80,
                        efficacy_grid = seq(0.95, 0.999, by = 0.001),
                        futility_grid = seq(0.80, 0.95,  by = 0.01))
print(cal$best)                     # calibrated design (adabay_calibration_best) at the targets

## Tidy-data view across all grid points (or a list of adabay_oc objects):
## summarise_oc(oc) returns a one-row data.frame; summarise_oc(list(...))
## stacks the three rows for inspection in tidyverse pipelines.
summarise_oc(list(oc, cal$best$oc_h0, cal$best$oc_h1))
\end{lstlisting}

The five output objects (\code{des}, \code{pri\_c}, \code{pri}, \code{dec}, \code{acc}) all provide print methods, and \code{pri\_c} additionally provides a plot method, allowing each stage to be inspected before the simulation pass. Step~5 evaluates a single uncalibrated design under $H_{1}$. Step~6 builds one cache per scenario over a shared threshold grid, then sweeps the posterior-probability thresholds against both caches to identify the design that satisfies the regulatory targets while minimising expected sample size. The optional \code{summarise\_oc()} helper combines one or more \code{adabay\_oc} objects into a tidy \code{data.frame} for reporting and downstream analysis. Replacing the binary endpoint specification with \code{endpoint = "continuous"}, \code{"count"} or \code{"tte"}, together with the corresponding effect-scale and family arguments, produces the case studies of Section~\ref{sec:5}.

\section{Worked examples}
\label{sec:5}
We present four case studies, one per endpoint type, each based on a published or representative trial. The four configurations are illustrative uncalibrated designs: the efficacy $p=0.99$ and futility thresholds $q=0.90$ are fixed at conventional Bayesian values rather than tuned to meet a regulatory target on the type I error rate. Their purpose is to exercise the user interface described in Section~\ref{sec:4} and to offer a basis for the software comparison in Section~\ref{sec:7}. The calibrated three- and five-look ADRENAL designs presented in Section~\ref{sec:63}, which use stage-specific thresholds $\boldsymbol{p}_{1:K}$ obtained by sweeping \code{evaluate\_design()} over a Haybittle--Peto-style schedule on the same cache, are the only configurations in this work that simultaneously meet the regulatory targets $\alpha\leq 2.5\%$ and $1-\beta\geq 90\%$. Each subsection below gives the \pkg{adabay} call sequence and the salient entries of the returned \code{adabay\_oc} object, and the full scripts are in the package's \code{inst/examples/} directory. Quantitative benchmarks against \pkg{gsbDesign}, \pkg{adaptr} and \pkg{BATSS} are postponed to Section~\ref{sec:7}.

\subsection{Continuous endpoint: the TRANSFORM-2 esketamine trial}
\label{sec:51}
TRANSFORM-2 \citep{popova2019} was a phase III, double-blind, active-controlled, randomised trial of flexibly dosed esketamine nasal spray plus a newly initiated oral antidepressant against placebo nasal spray plus a newly initiated oral antidepressant in adults with treatment-resistant depression. The primary endpoint was change from baseline in the Montgomery--{\AA}sberg Depression Rating Scale (MADRS, range $[0,60]$) total score at Day~28 of the four-week induction phase. The trial randomised $227$ patients in a $1{:}1$ ratio and reported a mean MADRS change of $-19.8$ in the esketamine arm versus $-15.8$ in the placebo arm, with a treatment-versus-control difference of $-4.0$ ($95\%$ confidence interval $-7.31$ to $-0.64$, $p=0.020$). The study contributed to the evidence base supporting the use of esketamine nasal spray for treatment-resistant depression.

We re-implement TRANSFORM-2 as a Bayesian GSD incorporating five looks ($K=5$) on the mean-difference scale. The data-generating model assumes $\mu_c=-15.8$ in the placebo arm and $\mu_t=-19.8$ in the esketamine arm, with a common within-arm standard deviation $\sigma=13$ for MADRS change, consistent with the per-arm dispersion reported in TRANSFORM-2. The clinically relevant effect is $\Delta^{*}=\mu_t-\mu_c=-4$, and the null is set to $\Delta_{H_0}=0$. For comparison, a fixed-sample design at $\alpha=0.025$ (one-sided) and $1-\beta=0.90$ requires $n=222$ patients per arm. The five-look GSD uses equally spaced looks at $n=45,89,133,177,222$ patients per arm. Both arms are assigned weakly informative priors $\mu_a\sim\mathcal{N}(-10,10^{6})$. The efficacy rule is the single criterion $\mathbb{P}(\Delta<0\mid\,\cdot\,)>0.99$ and the futility rule is $\mathbb{P}(\Delta>0\mid\,\cdot\,)>0.90$, both on the mean-difference scale.

The \pkg{adabay} call sequence is shown below. The four constructor calls create the design, prior and decision objects in Section~\ref{sec:4}. The two calls to \code{evaluate\_design()} then evaluate the operating characteristics under $H_0$ and $H_1$.

\begin{lstlisting}
des <- set_design(endpoint = "continuous",
                  n_per_look = c(90, 178, 266, 354, 444),
                  effect_scale = "mean_difference",
                  alternative = "less", sigma = 13)
pri <- set_prior(endpoint = "continuous",
                 arms = list(c = list(family = "normal", mean = -10, sd = 1e3),
                             t = list(family = "normal", mean = -10, sd = 1e3)))
dec <- set_decision(efficacy = list(list(threshold_effect = 0,
                                         threshold_prob   = 0.99)),
                    futility = list(list(threshold_effect = 0,
                                         threshold_prob   = 0.90)),
                    futility_binding = TRUE)
oc_h0 <- evaluate_design(des, pri, dec,
                         effect = list(mu_c = -15.8, mu_t = -15.8),
                         n_trials = 1e6, cores = 8, seed = 1L)
oc_h1 <- evaluate_design(des, pri, dec,
                         effect = list(mu_c = -15.8, mu_t = -19.8),
                         n_trials = 1e6, cores = 8, seed = 1L)
print(oc_h1)
plot(oc_h1)
\end{lstlisting}

At $R=10^{6}$ virtual trials per scenario, the re-design has a type I error rate of $3.09\%$ and power of $85.59\%$, with an expected sample size of $265$ patients across both arms under $H_1$ (versus the fixed-sample maximum of $444$). The full operating-characteristic comparison against \pkg{gsbDesign} (which integrates the analytical normal--normal posterior), \pkg{BATSS} and \pkg{adaptr} is reported in Table~\ref{tab:711} of Section~\ref{sec:7}. The \code{print()} method displays the per-look stopping probabilities, together with the overall efficacy and futility probabilities under the simulated effect. The \code{plot()} method draws a grouped bar chart of the per-look efficacy and futility stopping probabilities, showing how the trial accumulates decision mass across the five looks.

\subsection{Binary endpoint: the ADRENAL septic shock trial}
\label{sec:52}
The ADRENAL trial \citep{venkatesh2018} was a multicentre, placebo-controlled, double-blind, randomised clinical trial of hydrocortisone versus placebo in patients with septic shock. The primary endpoint was 90-day all-cause mortality. With a 33\% mortality rate in the placebo arm, the trial was powered to detect an absolute risk reduction of 5\% at a two-sided 5\% significance level, which required a maximum sample size of 3800 (1900 per arm). We re-implement the ADRENAL trial as a five-look Bayesian GSD ($K=5$) following \citet{he2026}, with looks equally spaced at $n=380,760,1140,1520,1900$ patients per arm, independent $\mathrm{Beta}(1,1)$ priors on the per-arm rates, a single-criterion efficacy rule $\mathbb{P}(\Delta<0\mid\,\cdot\,)>0.99$ on the risk-difference scale, and a futility rule $\mathbb{P}(\Delta>0\mid\,\cdot\,)>0.90$.

\begin{lstlisting}
des <- set_design(endpoint = "binary",
                  n_per_look = c(760, 1520, 2280, 3040, 3800),
                  effect_scale = "risk_difference",
                  alternative  = "less")
pri <- set_prior(endpoint = "binary",
                 arms = list(c = list(family = "beta", a = 1, b = 1),
                             t = list(family = "beta", a = 1, b = 1)))
dec <- set_decision(efficacy = list(list(threshold_effect = 0,
                                         threshold_prob   = 0.99)),
                    futility = list(list(threshold_effect = 0,
                                         threshold_prob   = 0.90)),
                    futility_binding = TRUE)
oc_h0 <- evaluate_design(des, pri, dec,
                         effect = list(theta_c = 0.33, theta_t = 0.33),
                         n_trials = 1e6, cores = 8, seed = 1L)
oc_h1 <- evaluate_design(des, pri, dec,
                         effect = list(theta_c = 0.33, theta_t = 0.28),
                         n_trials = 1e6, cores = 8, seed = 1L)
\end{lstlisting}

At $R=10^{6}$ virtual trials per scenario the uncalibrated design (common efficacy threshold $p=0.99$ and common futility threshold $q=0.90$ across looks) has a type I error rate of $3.12\%$ and a power of $87.74\%$, with an expected sample size of $\mathbb{E}(N\mid H_1)\approx 2202$ patients across both arms. The benchmarks against \pkg{BATSS} and \pkg{adaptr} are reported in Table~\ref{tab:721} of Section~\ref{sec:7}.

\subsection{Count endpoint: the Pfizer--BioNTech BNT162b2 vaccine trial}
\label{sec:53}
The Pfizer--BioNTech BNT162b2 trial \citep{polack2020} was a phase II/III randomised, observer-blinded, placebo-controlled vaccine-efficacy trial comparing two doses of the BNT162b2 mRNA vaccine with placebo saline in adults, for the prevention of laboratory-confirmed COVID-19 with onset at least seven days after the second dose. The primary analysis applied a Bayesian group sequential framework based on the cumulative number of confirmed cases across arms, with four planned interim looks at $32$, $62$, $92$ and $120$ accrued cases and a final analysis at $164$ cases. The protocol was designed to establish vaccine efficacy (VE) greater than $30\%$, corresponding to a rate ratio $\Delta^{*}=\lambda_t/\lambda_c<0.70$ under a design alternative of $60\%$, corresponding to a rate ratio of $0.40$. The assumed placebo-arm background incidence rate was approximately $\lambda_c=0.018$ cases per person-year. Under homogeneous Poisson event processes in the two arms, the accrued case counts follow the gamma--Poisson conjugate pair described in Section~\ref{sec:32}, so the Poisson count model implemented in \pkg{adabay} is a natural design-stage idealisation for this example.

We re-implement the BNT162b2 trial as a five-look Bayesian GSD ($K=5$) with looks placed at the cumulative person-time at which each prespecified case-count milestone is expected to occur under $H_1$. At the design alternative, $\bar\lambda=(\lambda_c+\lambda_t)/2\approx 0.0126$ cases per person-year, so the case-count milestones $32, 62, 92, 120, 164$ map to cumulative person-time values of approximately $E=1270, 2460, 3650, 4760, 6510$ per arm. We assign independent $\mathrm{Gamma}(1,1)$ priors on both $\lambda_c$ and $\lambda_t$.
The efficacy rule is the single criterion $\mathbb{P}(\log\Delta<\log 0.70\mid\,\cdot\,)>0.99$ on the log-rate-ratio scale, corresponding to the protocol superiority criterion for $\text{VE}>30\%$. The futility rule is $\mathbb{P}(\log\Delta>\log 0.70\mid\,\cdot\,)>0.90$.

The gamma--Poisson conjugate update in \pkg{adabay} reduces the posterior tail probability to a beta cumulative distribution function (see~\ref{apx:B}), so no quadrature is required and the per-trial cost is dominated by the count-data simulator.

\begin{lstlisting}
des <- set_design(endpoint          = "count",
                  exposure_per_look = c(2540, 4920, 7300, 9520, 13020),
                  effect_scale      = "log_rate_ratio",
                  alternative       = "less",
                  delta_null        = log(0.70))
pri <- set_prior(endpoint = "count",
                 arms = list(c = list(family = "gamma", a = 1, b = 1),
                             t = list(family = "gamma", a = 1, b = 1)))
dec <- set_decision(efficacy = list(list(threshold_effect = log(0.70),
                                         threshold_prob   = 0.99)),
                    futility = list(list(threshold_effect = log(0.70),
                                         threshold_prob   = 0.90)),
                    futility_binding = TRUE)
oc_h0 <- evaluate_design(des, pri, dec,
                         effect = list(lambda_c = 0.018, lambda_t = 0.0126),
                         n_trials = 1e6, cores = 8, seed = 1L)
oc_h1 <- evaluate_design(des, pri, dec,
                         effect = list(lambda_c = 0.018, lambda_t = 0.0072),
                         n_trials = 1e6, cores = 8, seed = 1L)
\end{lstlisting}

The $H_0$ scenario fixes the rate ratio at the superiority threshold ($\Delta^{*}=0.70$, $\text{VE}=30\%$), so the type I error rate reported under this scenario is the probability that the design declares $\text{VE}>30\%$ when the true $\text{VE}=30\%$. The $H_1$ scenario uses the design alternative ($\Delta^{*}=0.40$, $\text{VE}=60\%$). The observed trial outcome lay far inside the rejection region (eight vaccine cases versus 162 placebo cases at the final analysis, corresponding to $\text{VE}=95.0\%$). The trial crossed the protocol superiority boundary at the fourth interim. At $R=10^{6}$ virtual trials per scenario, the re-design has a type I error rate of $2.90\%$ and a power of $87.68\%$, with an expected cumulative person-time of $7662$ patient-years across both arms under $H_1$ (versus the maximum of $13020$ in the fixed-sample design). The benchmarks against \pkg{BATSS} are reported in Table~\ref{tab:722} of Section~\ref{sec:7}.

\subsection{Time-to-event endpoint: the CheckMate-141 head and neck cancer trial}
\label{sec:54}
CheckMate-141 \citep{ferris2016} was an open-label phase III randomised trial comparing nivolumab with investigator's choice of single-agent chemotherapy (methotrexate, docetaxel or cetuximab) in adults with recurrent or metastatic squamous-cell carcinoma of the head and neck (HNSCC) whose disease had progressed within six months after platinum-based chemotherapy. The primary endpoint was overall survival. The trial randomised $361$ patients in a $2{:}1$ allocation, accrued $218$ deaths by the planned interim analysis, at which it was terminated early for efficacy. Median overall survival was $7.5$ months with nivolumab versus $5.1$ months with chemotherapy, corresponding to a hazard ratio of $\Delta^{*}=\lambda_t/\lambda_c=0.70$ (one-decimal-place rounding of $5.1/7.5=0.68$; $97.73\%$ confidence interval $0.51$--$0.96$).

We re-implement CheckMate-141 as a five-look event-driven Bayesian GSD ($K=5$). For design-stage exposition, we adopt a $1{:}1$ allocation, noting that the actual $2{:}1$ randomisation can be accommodated by setting \code{allocation\_ratio = 2} in \code{set\_design()}. Interim analyses are triggered at cumulative event counts $D_k=50,100,150,200,250$. At the design alternative, the control and treatment hazards are set to $\lambda_c=\log 2/5.1=0.1359$ and $\lambda_t=\log 2/7.5=0.0924$ events per month, respectively. Weakly informative $\lambda_a\sim\mathrm{Gamma}(1,1)$ priors are applied to the arm-specific hazard rates. The efficacy rule is the single-criterion $\mathbb{P}(\log\Delta<0\mid\,\cdot\,)>0.99$ on the log-hazard-ratio scale, and the futility rule is $\mathbb{P}(\log\Delta>0\mid\,\cdot\,)>0.90$. Recruitment is modelled as a homogeneous Poisson process. CheckMate-141 randomised $361$ patients between 29 May 2014 and 31 July 2015 (about $14.1$ months), a pooled accrual rate of roughly $25.6$ patients per month, which \code{set\_accrual()} takes directly and splits by \code{allocation\_ratio} into $12.8$ patients per month per arm at the design-stage $1{:}1$ allocation.

The \pkg{adabay} call sequence shows the event-driven design and the recruitment specification; the expected calendar duration is obtained as the direct Monte Carlo mean of the realised stopping-look calendar time across the simulated trials.

\begin{lstlisting}
des <- set_design(endpoint     = "tte",
                  d_total      = 250,
                  d_per_look   = c(50, 100, 150, 200, 250),
                  effect_scale = "log_hazard_ratio",
                  alternative  = "less")
pri <- set_prior(endpoint = "tte",
                 arms = list(c = list(family = "gamma", a = 1, b = 1),
                             t = list(family = "gamma", a = 1, b = 1)))
dec <- set_decision(efficacy = list(list(threshold_effect = 0,
                                         threshold_prob   = 0.99)),
                    futility = list(list(threshold_effect = 0,
                                         threshold_prob   = 0.90)),
                    futility_binding = TRUE)
acc <- set_accrual(model = "poisson", rate = 25.6)
oc_h0 <- evaluate_design(des, pri, dec,
                         effect  = list(lambda_c = log(2)/5.1,
                                        lambda_t = log(2)/5.1),
                         accrual = acc,
                         n_trials = 1e6, cores = 8, seed = 1L)
oc_h1 <- evaluate_design(des, pri, dec,
                         effect  = list(lambda_c = log(2)/5.1,
                                        lambda_t = log(2)/7.5),
                         accrual = acc,
                         n_trials = 1e6, cores = 8, seed = 1L)
\end{lstlisting}

At $R=10^{6}$ virtual trials per scenario, the re-design has a type I error rate of $3.10\%$ and a power of $80.59\%$, with an expected number of events of $159$ at trial stop under $H_1$ (versus the fixed-sample target of $250$) and an expected calendar duration of $12.78$ months from randomisation of the first subject to the analysis at which the trial stops. No benchmark comparison is included for this example, because the comparator software considered in Section~\ref{sec:7} does not support this time-to-event setting.

\section{Calibration and caching}
\label{sec:6}
At the design stage, users typically examine operating characteristics across many candidate configurations of the number of looks $K$, the timing of looks $(\boldsymbol{n}_{c,1:K},\boldsymbol{n}_{t,1:K})$ and the threshold combination $(p_{k,u},q_{k,v})$. A naive implementation reruns the fused-path call mode of \code{evaluate\_design()} for each candidate, which performs a full Monte Carlo pass per evaluation and is wasteful: the per-look posterior tail probabilities of $\Delta$ do not depend on the thresholds, only on the data observed up to each look. \pkg{adabay} exposes this observation as a first-class software feature.

\subsection{Precomputation strategy}
\label{sec:61}
For a fixed simulation seed and a fixed endpoint type, the per-look posterior tail probabilities $\mathbb{P}(\Delta>e\mid\mathcal{D}_k)$ depend only on the data accumulated up to look $k$ and on the prior. The thresholds $(p_{k,u},q_{k,v})$ enter only through the indicator functions of Eq.~(\ref{eqn:313}). Once the trials have been simulated at the union of all candidate look times in the design grid, and the per-look tail probabilities have been cached at the union of all candidate effect-size thresholds $\{e_{k,u},f_{k,v}\}$, every candidate design configuration can be evaluated by a sweep over the cached tail probabilities and an aggregation over $\{(\tau^r,C^r)\}_{r=1}^{R}$. The expensive simulation pass is therefore performed exactly once, and any subsequent threshold or look-time configuration is evaluated at marginal cost.

\subsection{Building the cache}
\label{sec:62}
\code{build\_cache()} performs the one-time simulation pass and stores the per-look posterior tail probabilities at the union of all candidate effect-size thresholds $\{e_{k,u},f_{k,v}\}$. \code{evaluate\_design()} then scores any specific combination of posterior-probability thresholds $\{p_{k,u},q_{k,v}\}$ in milliseconds. The cache is an \code{adabay\_cache} S3 object with a \code{print()} method that displays the endpoint, the number of cached looks, the simulation budget and the union of effect-size thresholds.

\begin{lstlisting}
cache <- build_cache(
  design   = des,        # carries the union of candidate look times
  prior    = pri_conj,
  effect   = list(theta_c = 0.33, theta_t = 0.28),
  n_trials = 1e5,
  cores    = 8,
  seed     = 1L,
  threshold_grid = list( # effect-size thresholds e and f of Eqs. (1)-(2)
    efficacy = c(0.0, -0.05),
    futility = 0.0
  )
)

## Two scoring sweeps over the same cache, at two posterior-probability
## threshold combinations (p, q of Eqs. (1)-(2)), each completing in
## milliseconds rather than minutes
dec_a <- set_decision(
  efficacy = list(list(threshold_effect = 0, threshold_prob = 0.99)),
  futility = list(list(threshold_effect = 0, threshold_prob = 0.90)),
  futility_binding = FALSE)
dec_b <- set_decision(
  efficacy = list(list(threshold_effect = 0, threshold_prob = 0.995)),
  futility = list(list(threshold_effect = 0, threshold_prob = 0.95)),
  futility_binding = FALSE)
oc_a <- evaluate_design(cache, dec_a)  # p = 0.99,  q = 0.90
oc_b <- evaluate_design(cache, dec_b)  # p = 0.995, q = 0.95
\end{lstlisting}

\subsection{Threshold calibration}
\label{sec:63}
\code{calibrate\_design()} is the convenience wrapper that takes a pair of caches (one cached under $H_{0}$, one under $H_{1}$), searches a user-supplied grid of posterior-probability thresholds $(p,q)$ under target $(\alpha,1-\beta)$ constraints, and returns the design that meets the targets while minimising the expected sample size. The two-cache requirement is intentional. The type I error rate is read off the cache under $H_{0}$, and the power and the expected sample size are read off the cache under $H_{1}$. A single cache therefore cannot in general supply both quantities.

\begin{lstlisting}
cache_h0 <- build_cache(design = des, prior = pri_conj,
                        effect = list(theta_c = 0.33, theta_t = 0.33),
                        n_trials = 1e5, cores = 8, seed = 1L,
                        threshold_grid = list(efficacy = 0, futility = 0))
cache_h1 <- build_cache(design = des, prior = pri_conj,
                        effect = list(theta_c = 0.33, theta_t = 0.28),
                        n_trials = 1e5, cores = 8, seed = 1L,
                        threshold_grid = list(efficacy = 0, futility = 0))
cal <- calibrate_design(
  cache         = cache_h0,
  cache_alt     = cache_h1,
  alpha_target  = 0.025,
  power_target  = 0.85,
  efficacy_grid = seq(0.95, 0.999, by = 0.001), # grid of p values
  futility_grid = seq(0.80, 0.95,  by = 0.01)   # grid of q values
)
print(cal$best)        # calibrated thresholds, achieved alpha and power
cal$best$expected_sample_size  # expected sample size at the calibrated thresholds
\end{lstlisting}

The power target is set to $0.85$ here because no single common-across-looks threshold pair on this five-look design attains $90\%$ power at $\alpha\le 2.5\%$: over the grid above, the achievable power subject to the type I error constraint peaks at $86.08\%$, attained at $(p,q)=(0.992,0.80)$ with a type I error rate of $2.43\%$ and $\mathbb{E}(N\mid H_1)\approx 2272$. The sweep runs under \code{calibrate\_design()}'s default non-binding futility rule, whereas the stage-specific designs below are evaluated under binding futility, so the two are not on a common footing; the sweep is reported only to show that no common threshold pair reaches the power target. This is precisely the limitation that motivates the stage-specific schedules considered next.

The binary ADRENAL re-design is the main example in the companion methodology paper \citep{he2026}, which evaluates designs with up to nine equally spaced interim analyses and presents a 438-design calibration grid. The current implementation of \code{calibrate\_design()} sweeps a single common-across-looks threshold. Designs with stage-specific thresholds, including the Haybittle--Peto-style schedules with stringent interim thresholds and a near-nominal final threshold used in the companion paper, are therefore evaluated from the same cache using a short user-defined loop that calls \code{evaluate\_design()} with the stage-specific decision rule (a list of $K$ per-look criteria passed to \code{set\_decision()}). We illustrate the loop on two designs of that family. The five-look design with stage-specific efficacy thresholds $\boldsymbol{p}_{1:5}=(0.997,0.997,0.997,0.997,0.980)$ and a common futility threshold $q=0.85$ is formally distinct from the uncalibrated $(p=0.99, q=0.90)$ ADRENAL design of Section~\ref{sec:52} (Table~\ref{tab:721}). It has type I error rate of $2.42\%$, power of $90.18\%$, $\mathbb{E}(N\mid H_0)\approx 3071$ and $\mathbb{E}(N\mid H_1)\approx 2638$. The corresponding three-look design at $\boldsymbol{p}_{1:3}=(0.995,0.995,0.980)$ and the same $q=0.85$ has type I error rate of $2.49\%$, power of $90.54\%$, $\mathbb{E}(N\mid H_0)\approx 3315$ and $\mathbb{E}(N\mid H_1)\approx 2725$. These threshold schedules are our own: they are not the schedules calibrated in \citet{he2026}, whose maximum-power frontier holds the interim threshold at $0.999$ with a near-nominal final threshold and a common futility threshold $q=0.90$. The designs here exercise the stage-specific evaluation path rather than reproduce that calibration. Both designs are evaluated by the same loop: the five-look design from the cache above, and the three-look design from a companion three-look cache built in the same way on three equally spaced looks. A cache stores per-look tail probabilities at a fixed set of look times, and any design whose looks form a subset of that set can be scored from the same cache through the \code{look\_subset} argument of the cached path; the three equally spaced looks used here are not a subset of the five-look schedule, so this design requires its own cache. A future minor release will extend \code{calibrate\_design()} to search a Cartesian product of per-look threshold grids natively.

\subsection{Direct simulation versus cached grid evaluation}
\label{sec:64}
Table~\ref{tab:641} compares direct simulation with the caching function for the ADRENAL re-design (Section~\ref{sec:52}) at $R=10^{5}$. A grid of $50\times 16=800$ threshold combinations is evaluated through two approaches. The direct approach reruns the fused-path \code{evaluate\_design()} call (on an \code{adabay\_design}) for each combination, whereas the cached approach performs a single \code{build\_cache()} pass and then evaluates the grid by repeated cached-path \code{evaluate\_design()} calls (on the resulting \code{adabay\_cache}). Therefore, the cached approach amortises a single simulation pass across the full grid, while direct simulation incurs the simulation cost $800$ times. The numbers in Table~\ref{tab:641} are reproduced by the Section~6.4 block of the replication script \code{replication.R} distributed with the package, and are consistent with the $438$-design grid calibration of \citet{he2026}, which amortises a single simulation pass (a one-time cost of approximately $30$~seconds at $R=10^{6}$ on eight logical threads) over the whole grid at about $0.6$~seconds per additional design.

\begin{table}[!ht]
\centering
\begin{tabular}{lcc}
\toprule
Step & Direct simulation (s) & Cached grid evaluation (s) \\
\midrule
One-time simulation pass                 & not applicable & $6.8$ \\
Per-design score (mean over $800$)       & $6.9$           & $0.034$ \\
Total wall-clock for the $800$-cell grid & ${\sim}5500$ & ${\sim}34$ \\
\bottomrule
\end{tabular}
\caption{Direct simulation versus cached grid evaluation for the ADRENAL re-design at $R=10^{5}$, on $8$ CPU cores. The one-time simulation pass cost of $6.8$~s is the measured wall-clock of a single \code{build\_cache()} call at $R=10^{5}$ on $8$ cores; cache building stores per-look posterior tail probabilities at every grid threshold, at a per-trial cost comparable to a single fused-path \code{evaluate\_design()} call, so $6.8$~s at $R=10^{5}$ scales linearly to ${\sim}68$~s at $R=10^{6}$ (compared with $137.9$~s for the two-scenario fused-path pass in Table~\ref{tab:721}). The cached path performs the simulation pass exactly once and amortises it across the grid, so only the one-time pass scales with $R$, whereas the direct path repeats the pass for every threshold combination.}
\label{tab:641}
\end{table}

\subsection{Memory footprint}
\label{sec:65}
The main memory requirement of the caching function is the \code{adabay\_cache} object, which stores the per-look posterior tail probabilities for every virtual trial at the union of all candidate effect-size thresholds. Let $N_T$ denote the size of this union, that is, the number of unique effect-size thresholds at which tail probabilities are cached. For $R$ virtual trials, $K$ candidate look times in the union, and $N_T$ unique effect-size thresholds, the cache holds approximately $R K N_T$ double-precision floats, that is, $8 R K N_T$ bytes. 

For the ADRENAL grid in Table~\ref{tab:641} ($R=10^{5}$ virtual trials, $K=5$ look times and the single effect-size threshold $N_T=1$ shared by the efficacy and futility rules), this is approximately $4$~MB ($8\cdot 10^{5}\cdot 5\cdot 1$ bytes). The same cache supports any combination of posterior-probability thresholds $\{p_{k,u},q_{k,v}\}$ at no additional storage cost. For the $438$-design grid of \citet{he2026} (with $K$ up to $9$, the union of all candidate look times of size up to $45$ and $N_T=3$ effect-size thresholds), the cache scales to roughly $11$~MB at $R=10^{4}$ and to about $1.1$~GB at $R=10^{6}$, both well within the working memory of a commodity workstation. 

Direct simulation has a smaller memory footprint since it only needs to retain the per-trial result vector after each pass, therefore using memory of order $R$ per scenario. However, it pays the full simulation cost separately for every threshold combination.

\section{Validation and benchmarking}
\label{sec:7}
In this section, we validate and benchmark \pkg{adabay} against the closest available software. The comparators are \pkg{gsbDesign} \citep{gerber2016,gerber2024}, which provides analytical normal--normal updates for continuous endpoints; \pkg{adaptr} \citep{granholm2022,granholm2026}, which uses closed-form conjugate updates for Gaussian and binomial outcomes; and \pkg{BATSS} \citep{couturier2024,couturier2025}, which implements INLA-based GLMs for continuous, binary and count endpoints but does not currently support time-to-event outcomes. \pkg{adabay} is evaluated at two simulation budgets: $R=10^{4}$ virtual trials per scenario for a matched-budget comparison with the simulation-based comparators, and $R=10^{6}$ virtual trials per scenario as a high-precision reference. \pkg{BATSS} and \pkg{adaptr} are run at the matched budget $R=10^{4}$, while \pkg{gsbDesign}, the only comparator that returns operating characteristics analytically rather than by simulation, is run in its native numerical-integration mode in the continuous-endpoint case study.

\pkg{adabay} is run on $8$ cores via \code{mclapply} with the L'Ecuyer combined multiple-recursive generator described in Section~\ref{sec:46} (\code{cores = 8}), matching the eight logical threads of the reference workstation. \pkg{adaptr} is run on $8$ cores via its native parallel backend (\code{cores = 8}). \pkg{gsbDesign} is run on a single core (native single-threaded numerical integration). We run \pkg{BATSS} with eight workers via \code{mclapply}, with \pkg{INLA} pinned to one thread per worker (\code{num.threads = "1:1"}). The wall-clock times reported in Tables~\ref{tab:711}--\ref{tab:723} therefore correspond to these package-specific configurations. For a like-for-like per-virtual-trial comparison, all simulation-based packages are run on $8$ cores at the matched budget $R=10^{4}$. The wall-clock times in Tables~\ref{tab:711}--\ref{tab:723} are therefore directly comparable at equal parallelism (see Section~\ref{sec:74}).

The benchmark design is a five-look Bayesian GSD ($K=5$) with looks equally spaced across the maximum information of the corresponding fixed-sample design. It adopts a single-criterion efficacy rule with $p=0.99$ and a futility rule with $q=0.90$, and is evaluated at both the global null hypothesis $H_0:\Delta=\Delta_{H_0}$ and the alternative anchored at the clinically relevant effect $H_1:\Delta=\Delta^{*}$. All benchmarks in this section use binding futility (\code{futility\_binding = TRUE} in the case-study scripts of Section~\ref{sec:5}). The corresponding non-binding type~I error rate, defined in Section~\ref{sec:31} as the rate that would be observed if the futility rule were ignored, is the regulatory default and is typically a few tenths of a percentage point higher. It can be obtained from the same simulation streams by re-evaluating the cached per-look tail probabilities of Section~\ref{sec:6} with the futility rule suppressed.

\subsection{Agreement with \pkg{gsbDesign}}
\label{sec:71}
For the continuous endpoint of Section~\ref{sec:51}, \pkg{gsbDesign} provides an exact reference by computing the operating characteristics analytically through numerical integration of the normal--normal conjugate posterior. Table~\ref{tab:711} reports the \pkg{adabay} estimates at the matched budget and at $R=10^{6}$, alongside the \pkg{BATSS} and \pkg{adaptr} estimates at the matched budget and the \pkg{gsbDesign} analytic values. At $R=10^{6}$, the \pkg{adabay} Monte Carlo estimates agree with the \pkg{gsbDesign} analytic values to within $0.05$ percentage points on the type~I error rate and the power and to within half a patient on the expected sample sizes, serving as an exact-reference cross-check of the continuous-endpoint implementation. The \pkg{BATSS} estimates at the matched budget are consistent with the \pkg{adabay} estimates within Monte Carlo error, as are the \pkg{adaptr} power and expected sample sizes. The one entry outside two Monte Carlo standard errors is the \pkg{adaptr} type~I error rate of $3.60\%$, which lies $0.51$ percentage points above the \pkg{gsbDesign} analytic value of $3.09\%$. The standard errors in Table~\ref{tab:711} reflect trial-level binomial variability only, and do not account for the additional error that \pkg{adaptr} incurs by evaluating its posterior probabilities from a finite set of posterior draws against the sharp $0.99$ efficacy threshold at each look, whereas \pkg{adabay} and \pkg{gsbDesign} evaluate the corresponding tail probabilities exactly. Allowing for this residual, the three independent simulation-based packages and the analytic reference lead to concordant operating characteristics for the continuous-endpoint case study.

\begin{table}[!ht]
\centering
\resizebox{\textwidth}{!}{%
\begin{tabular}{lcccccc}
\toprule
Method & \code{n\_trials} & Type I (MC SE) & Power (MC SE) & $\mathbb{E}(N\mid H_0)$ & $\mathbb{E}(N\mid H_1)$ & Time (s) \\
\midrule
\pkg{adabay}    & $1{,}000{,}000$ & 3.09\,(0.02) & 85.59\,(0.04) & 379 & 265 &   14.6 \\
\pkg{adabay}    &  $10{,}000$ & 2.92\,(0.17) & 86.18\,(0.35) & 381 & 265 &    0.2 \\
\pkg{BATSS}     &  $10{,}000$ & 3.23\,(0.18) & 85.46\,(0.35) & 381 & 264 & ${\sim}30{,}724$ \\
\pkg{adaptr}    &  $10{,}000$ & 3.60\,(0.19) & 85.40\,(0.35) & 379 & 264 &   66.0 \\
\pkg{gsbDesign} &    analytic & 3.09\,($-$)  & 85.54\,($-$)  & 379 & 266 &    0.3 \\
\bottomrule
\end{tabular}%
}
\caption{Operating characteristics of the TRANSFORM-2 Bayesian GSD re-design (Section~\ref{sec:51}, continuous endpoint) at $K=5$, $p=0.99$ and $q=0.90$. The column \code{n\_trials} is the Monte Carlo simulation budget per scenario. \pkg{gsbDesign} computes the operating characteristics by numerical integration of the analytical normal--normal posterior and is listed as ``analytic''; being deterministic, it carries no Monte Carlo standard error, shown as ``$-$''. \pkg{adabay} runs on $8$ CPU cores via \code{mclapply} with the L'Ecuyer combined multiple-recursive generator of Section~\ref{sec:46}. \pkg{adaptr} runs on $8$ CPU cores via its native parallel backend; \pkg{gsbDesign} runs on a single CPU core. \pkg{BATSS} runs with eight \code{mclapply} workers and \pkg{INLA} pinned to one thread per worker. ``MC SE'' is the Monte Carlo standard error in percentage points. The fixed-sample maximum is $N_{\max}=444$ across both arms. Type I error is evaluated at $\mu_t-\mu_c=0$ and power at $\mu_t-\mu_c=-4$ (the observed TRANSFORM-2 mean-difference effect). The Time column is the sum of the $H_0$ and $H_1$ wall-clock costs. The four packages adopt different default priors: \pkg{adabay} uses independent normal priors $\mu_a\sim\mathcal{N}(-10,10^{6})$ on each arm, \pkg{gsbDesign} uses its non-informative prior on the treatment difference, \pkg{BATSS} uses its default \pkg{INLA} fit with weakly informative Gaussian priors on the regression coefficients, and \pkg{adaptr} uses its default flat improper prior on the treatment difference.}
\label{tab:711}
\end{table}

\subsection{Agreement with simulation-based packages \pkg{BATSS} and \pkg{adaptr}}
\label{sec:72}
For the binary endpoint of Section~\ref{sec:52}, Table~\ref{tab:721} compares \pkg{adabay} to \pkg{BATSS} (binomial family with logit link) and \pkg{adaptr}. For the count endpoint of Section~\ref{sec:53}, Table~\ref{tab:722} compares \pkg{adabay} with \pkg{BATSS}. In this scenario, the \pkg{BATSS} analysis uses the effect-margin arguments \code{delta.eff} and \code{delta.fut} in \code{batss.glm}, both set to $\log(0.70)$ (the protocol threshold $\text{VE}>30\%$). The null scenario is simulated at the protocol rate ratio $0.70$ rather than \pkg{BATSS}'s default no-effect ratio of $1$. Since \pkg{BATSS} does not directly implement exposure-driven interim schedules, the comparison applies a residual person-time-proxy approximation to the cumulative person-time analysis schedule. Under this approximation, the \pkg{BATSS} estimates agree with the \pkg{adabay} estimates within Monte Carlo error. For the time-to-event endpoint of Section~\ref{sec:54}, Table~\ref{tab:723} reports only the \pkg{adabay} estimates, because neither \pkg{adaptr} nor \pkg{BATSS} supports the required endpoint type.

\begin{table}[!ht]
\centering
\resizebox{\textwidth}{!}{%
\begin{tabular}{lcccccc}
\toprule
Method & \code{n\_trials} & Type I (MC SE) & Power (MC SE) & $\mathbb{E}(N\mid H_0)$ & $\mathbb{E}(N\mid H_1)$ & Time (s) \\
\midrule
\pkg{adabay} & $1{,}000{,}000$ & 3.12\,(0.02) & 87.74\,(0.03) & 3{,}241 & 2{,}202 & 137.9 \\
\pkg{adabay} &  $10{,}000$ & 2.88\,(0.17) & 88.15\,(0.32) & 3{,}251 & 2{,}200 &   1.6 \\
\pkg{BATSS}  &  $10{,}000$ & 2.95\,(0.17) & 87.97\,(0.33) & 3{,}240 & 2{,}196 & ${\sim}32{,}317$ \\
\pkg{adaptr} &  $10{,}000$ & 3.40\,(0.18) & 87.50\,(0.33) & 3{,}234 & 2{,}206 &  62.7 \\
\pkg{gsbDesign} & \textsc{n/a} & \textsc{n/a} & \textsc{n/a} & \textsc{n/a} & \textsc{n/a} & \textsc{n/a} \\
\bottomrule
\end{tabular}%
}
\caption{Operating characteristics of the ADRENAL Bayesian GSD re-design (Section~\ref{sec:52}, binary endpoint) at $K=5$, $p=0.99$ and $q=0.90$. Computational conventions and the \code{n\_trials} and ``MC SE'' columns are as in Table~\ref{tab:711}. The fixed-sample maximum is $N_{\max}=3800$ across both arms. Type I error is evaluated at $\vartheta_t-\vartheta_c=0$ and power at $\vartheta_t-\vartheta_c=-0.05$. Priors are independent $\mathrm{Beta}(1,1)$ in each arm for \pkg{adabay} and \pkg{adaptr}, whereas \pkg{BATSS} uses its default \pkg{INLA} fit with weakly informative Gaussian priors on the logit-scale regression coefficients. The \pkg{adaptr} and \pkg{BATSS} estimates both agree with \pkg{adabay} within Monte Carlo error at the matched budget $R=10^{4}$. ``n/a'' indicates that the package does not support binary endpoints: \pkg{gsbDesign} is restricted to the normal--normal continuous setting.}
\label{tab:721}
\end{table}

\begin{table}[!ht]
\centering
\resizebox{\textwidth}{!}{%
\begin{tabular}{lcccccc}
\toprule
Method & \code{n\_trials} & Type I (MC SE) & Power (MC SE) & $\mathbb{E}(E\mid H_0)$ & $\mathbb{E}(E\mid H_1)$ & Time (s) \\
\midrule
\pkg{adabay} & $1{,}000{,}000$ & 2.90\,(0.02) & 87.68\,(0.03) & 11{,}046 & 7{,}662 &   4.7 \\
\pkg{adabay} &  $10{,}000$ & 3.02\,(0.17) & 88.37\,(0.32) & 11{,}038 & 7{,}636 &   0.1 \\
\pkg{BATSS}  &  $10{,}000$ & 3.04\,(0.17) & 87.91\,(0.33) & 11{,}031 & 7{,}610 & ${\sim}35{,}776$ \\
\pkg{adaptr} & \textsc{n/a} & \textsc{n/a} & \textsc{n/a} & \textsc{n/a} & \textsc{n/a} & \textsc{n/a} \\
\pkg{gsbDesign} & \textsc{n/a} & \textsc{n/a} & \textsc{n/a} & \textsc{n/a} & \textsc{n/a} & \textsc{n/a} \\
\bottomrule
\end{tabular}%
}
\caption{Operating characteristics of the BNT162b2 COVID-19 vaccine Bayesian GSD re-design (Section~\ref{sec:53}, count endpoint) at $K=5$, $p=0.99$ and $q=0.90$. Computational conventions and the \code{n\_trials} and ``MC SE'' columns are as in Table~\ref{tab:711}. The \pkg{BATSS} comparison is constructed with \code{batss.glm}'s \code{delta.eff}/\code{delta.fut} margin set to $\log(0.70)$ and the null simulated at the protocol rate ratio $0.70$ (Section~\ref{sec:72}); it agrees with the \pkg{adabay} values within Monte Carlo error, under a residual person-time-proxy approximation that maps \pkg{BATSS}'s recruited-patient interims onto the exposure-driven cumulative-person-time look schedule.
The fixed-sample maximum is $13020$ person-years across both arms. Type I error is evaluated at $\lambda_t/\lambda_c=0.70$ (the protocol superiority threshold corresponding to $\text{VE}=30\%$) and power at $\lambda_t/\lambda_c=0.40$ (the protocol designed-to-detect alternative corresponding to $\text{VE}=60\%$). Priors are independent $\mathrm{Gamma}(1,1)$ in each arm for \pkg{adabay},
whereas \pkg{BATSS} uses its default \pkg{INLA} fit with weakly informative Gaussian priors on the log-rate regression coefficients. ``n/a'' indicates that the package does not support count endpoints: \pkg{adaptr} covers only continuous and binary outcomes, and \pkg{gsbDesign} is restricted to the normal--normal continuous setting.}
\label{tab:722}
\end{table}

\begin{table}[!ht]
\centering
\resizebox{\textwidth}{!}{%
\begin{tabular}{lcccccc}
\toprule
Method & \code{n\_trials} & Type I (MC SE) & Power (MC SE) & $\mathbb{E}(D\mid H_1)$ & $\mathbb{E}(T\mid H_1)$ & Time (s) \\
\midrule
\pkg{adabay} & $1{,}000{,}000$ & 3.10\,(0.02) & 80.59\,(0.04) & 159 & 12.78 &   75.1 \\
\pkg{adabay} &  $10{,}000$ & 3.29\,(0.18) & 80.56\,(0.40) & 158 & 12.73 &    0.7 \\
\pkg{BATSS}  & \textsc{n/a} & \textsc{n/a} & \textsc{n/a} & \textsc{n/a} & \textsc{n/a} & \textsc{n/a} \\
\pkg{adaptr} & \textsc{n/a} & \textsc{n/a} & \textsc{n/a} & \textsc{n/a} & \textsc{n/a} & \textsc{n/a} \\
\pkg{gsbDesign} & \textsc{n/a} & \textsc{n/a} & \textsc{n/a} & \textsc{n/a} & \textsc{n/a} & \textsc{n/a} \\
\bottomrule
\end{tabular}%
}
\caption{Operating characteristics of the CheckMate-141 HNSCC Bayesian GSD re-design (Section~\ref{sec:54}, time-to-event endpoint) at $K=5$, $p=0.99$ and $q=0.90$. Computational conventions and the \code{n\_trials} and ``MC SE'' columns are as in Table~\ref{tab:711}. ``n/a'' indicates that the package does not support time-to-event endpoints: \pkg{BATSS}'s sole model interface \code{batss.glm} is GLM-only and cannot construct the censored event-time response that an \pkg{INLA} survival likelihood requires (a \pkg{BATSS}-interface limitation, not an \pkg{INLA} one), with the event-driven look schedule a further obstacle (Section~\ref{sec:72}); \pkg{adaptr} covers only continuous and binary outcomes; and \pkg{gsbDesign} is restricted to the normal--normal continuous setting. Type I error is evaluated at $\lambda_t/\lambda_c=1$ (equal median overall survival of $5.1$ months in both arms) and power at $\lambda_t/\lambda_c=0.68$ (control median $5.1$ months versus treatment median $7.5$ months, the CheckMate-141 hazard-ratio target of $0.70$ to one decimal place). $\mathbb{E}(D\mid H_1)$ is the expected number of events at trial stop, and $\mathbb{E}(T\mid H_1)$ is the expected calendar duration from randomisation of the first subject to the analysis at which the trial stops, in months. It includes both the recruitment time and the follow-up time required to observe the trigger event count, and is obtained for \pkg{adabay} as the direct Monte Carlo mean of the realised stopping-look calendar time under Poisson recruitment at a pooled rate of $25.6$ patients per month ($12.8$ per arm at the $1{:}1$ allocation).}
\label{tab:723}
\end{table}

The \pkg{adabay} estimates are reported at two simulation budgets: $R=10^{4}$, the matched budget for the simulation-based comparators, and $R=10^{6}$, the high-precision reference. For the binary ADRENAL re-design, at the matched budget, the \pkg{adaptr} and \pkg{BATSS} estimates both agree with the \pkg{adabay} values within Monte Carlo error and reproduce the $R=5000$ binary comparison of \citet{he2026}. For the count BNT162b2 re-design, the \pkg{BATSS} estimates are similarly consistent with the high-precision \pkg{adabay} values within Monte Carlo error,
under a recruited-patient person-time proxy for the exposure-driven cumulative-person-time look schedule.

\subsection{Scalability}
\label{sec:73}
The benchmark tables in Sections~\ref{sec:71} and~\ref{sec:72} report a single design configuration at a fixed simulation budget. We now assess how \pkg{adabay} scales along the three dimensions that drive design-stage computational cost: the simulation size $R$, the number of looks $K$ and the size of the design grid $N_D$, where $N_D$ denotes the number of candidate design configurations evaluated during calibration.

\paragraph{Scaling with $R$} The trial-level loop in Algorithm~\ref{alg:351} is embarrassingly parallel, and the per-trial cost is constant for fixed $K$ and endpoint, so wall-clock time scales linearly in $R$ at fixed worker count. The Monte Carlo standard error of each operating-characteristic estimator decreases at the usual $1/\sqrt{R}$ rate. \citet{fda2010,fda2026} expect threshold calibration to be supported by simulation studies that estimate the type~I error rate with precision well below the target rate itself. For a one-sided target of $\alpha=2.5\%$, this translates to a Monte Carlo standard error on the order of $0.05$ percentage points, requiring approximately $R \geq 10^{5}$ virtual trials per scenario. \pkg{adabay} completes such budgets in seconds on commodity hardware.

\paragraph{Scaling with $K$} Per-look computation is dominated by the conjugacy-preserving posterior update and the one-dimensional tail-probability evaluation of Table~\ref{tab:321}, both of which are constant per look. Wall-clock time therefore scales approximately linearly with $K$. The variance of the type I error rate estimator increases weakly with $K$ because additional interim looks inflate the cumulative early stopping probability under $H_0$. In the binary benchmark of \citet{he2026}, the Monte Carlo standard error increases from $0.31$ to $0.64$ percentage points as $K$ rises from $1$ to $9$ at $R=1000$; the ratio across $K$ is preserved at every $R$ because the per-look posterior update is independent of $R$.

\paragraph{Scaling with $N_D$} Direct evaluation of $N_D$ candidate design configurations, defined by combinations of look times and posterior-probability thresholds, requires $N_D$ separate simulation passes and therefore scales linearly with $N_D$. The caching architecture in Section~\ref{sec:6} amortises a single simulation pass over the full grid, so the marginal cost of an additional configuration is the per-design score of $0.03$ seconds reported in Table~\ref{tab:641}. This decoupling is the main practical advantage of the package during calibration. For example, \citet{he2026} calibrate a $438$-design grid from a single simulation pass per hypothesis, reported as a one-time cost of approximately $30$~seconds at $R=10^{6}$ on eight logical threads, after which each additional design adds about $0.6$~seconds, so the full grid is scored in roughly four minutes per hypothesis. The corresponding direct approach would require $438$ separate \pkg{BATSS} or \pkg{adaptr} runs at full per-scenario cost, which \citet{he2026} put at roughly a CPU-month per hypothesis under \pkg{BATSS} on the same hardware.

\subsection{Runtime}
\label{sec:74}
The Time columns of Tables~\ref{tab:711}--\ref{tab:723} report the wall-clock cost at the matched budget convention $R=10^{4}$. For \pkg{adabay}, run on $8$ cores, the paired $H_0$ and $H_1$ scenarios complete in $0.1$ to $1.6$ seconds across the four endpoints (the Time column is the sum of the two scenario wall-clocks). 
\pkg{BATSS}, run with eight \code{mclapply} workers and \pkg{INLA} pinned to one thread per worker, completes the paired $H_0$ and $H_1$ scenarios in approximately $30724$ seconds ($\approx 8.5$ hours) in the continuous case, $32317$ seconds ($\approx 9$ hours) in the binary case, and $35776$ seconds ($\approx 9.9$ hours) in the count case at $R=10^{4}$.
\pkg{adaptr}, also run on $8$ cores, completes the paired scenarios in $66.0$ seconds in the continuous case and $62.7$ seconds in the binary case at $R=10^{4}$. The time-to-event comparisons with \pkg{BATSS} and \pkg{adaptr} are not constructible, as discussed in Section~\ref{sec:72}, so no comparator runtime is reported for that endpoint.

All three simulation-based packages are run on $8$ CPU cores at the matched budget $R=10^{4}$, so their wall-clock times are directly comparable at equal parallelism and an equal simulation budget. The per-virtual-trial speedup is then simply the ratio of the paired-scenario wall clocks. \pkg{gsbDesign} is excluded from this comparison since it returns operating characteristics by deterministic numerical integration on a single core rather than by simulation, and is sub-second. For the binary ADRENAL re-design at $R=10^{4}$, \pkg{adabay} completes the paired scenarios in $1.6$ seconds, \pkg{adaptr} in $62.7$ seconds and \pkg{BATSS} in $32317$ seconds, all on $8$ cores. The resulting per-virtual-trial speedups of \pkg{adabay} are approximately $39\times$ over \pkg{adaptr} (between one and two orders of magnitude) and approximately $2.0\times10^{4}$ over \pkg{BATSS} (four orders of magnitude), reproducing under the present $R=10^{4}$ convention the binary-endpoint result of \citet{he2026}. The same ordering holds for the continuous endpoint ($0.2$ seconds for \pkg{adabay} versus $66.0$ seconds for \pkg{adaptr} and $30724$ seconds for \pkg{BATSS}, all on $8$ cores) and for the count endpoint ($0.1$ seconds for \pkg{adabay} versus $35776$ seconds for \pkg{BATSS}). 

At the high-precision reference budget $R=10^{6}$, \pkg{adabay}, run on 8 cores, completes the paired $H_0$ and $H_1$ scenarios in $5$ to $138$ seconds across the four endpoints, as demonstrated in the Time columns of Tables~\ref{tab:711}--\ref{tab:723}. The corresponding Monte Carlo standard errors for the type I error rate and power are between $0.02$ to $0.04$ percentage points. The \pkg{gsbDesign} analytic path for the continuous case is faster still, with sub-second runtime, at the cost of being restricted to the normal--normal setting.

\subsection{Validation suite}
\label{sec:75}
\pkg{adabay} is validated through four complementary checks. First, the \pkg{testthat}-based unit-test suite described in Section~\ref{sec:46} is run on every package install and covers the constructors, mixture fitter, per-look posterior computation, simulator and caching function. Second, release-level regression tests compare fixed reference vectors of per-look stopping probabilities and overall $(\alpha,1-\beta)$ values generated from frozen seeds for the four case studies in Section~\ref{sec:5}. Any numerical deviation triggers a build failure. Third, analytic and cross-package checks provide independent external validation. For the continuous case, \pkg{adabay} agrees with the deterministic normal--normal reference from \pkg{gsbDesign} to within 0.05 percentage points. It is also consistent within Monte Carlo error with \pkg{adaptr} for the binary case and, apart from the type~I error rate discussed in Section~\ref{sec:71}, for the continuous case, and with \pkg{BATSS} for the binary, continuous and count cases. Fourth, reproducibility under parallel execution is assessed through the L'Ecuyer combined multiple-recursive generator described in Section~\ref{sec:46}. For any fixed \code{(seed, cores)} pair, the per-trial outputs and returned \code{adabay\_oc} object are bit-identical across reruns, which is verified by rerunning the ADRENAL re-design twice at \code{cores = 4} and twice at \code{cores = 8}, for both $R=10^{4}$ and $R=10^{6}$. Within each \code{cores} setting, the per-look stopping probabilities and overall $(\alpha,1-\beta)$ values match bit-for-bit. The returned objects differ only in their wall-clock timestamps. With \code{cross\_core\_reproducible = FALSE} (the default), running the same design at different worker counts produces operating characteristics that agree within Monte Carlo error but are not bit-identical, because the simulator's chunk partition changes with the number of workers. Setting \code{cross\_core\_reproducible = TRUE} on \code{evaluate\_design()} and \code{build\_cache()} (Section~\ref{sec:46}) restores bit-identity across worker counts by forcing the simulator pass into a single driver process; this is verified by a regression test that compares the returned \code{adabay\_oc} objects from \code{cores = 1}, \code{cores = 2} and \code{cores = 4} on a three-look binary design and confirms exact equality of the overall type I error rate, the overall efficacy probability, the per-look efficacy and futility stopping probabilities and the expected sample size.

\subsection{Feature comparison}
\label{sec:76}
Table~\ref{tab:761} compares \pkg{adabay} with the closest existing tools across four feature axes: endpoint coverage, flexible non-conjugate priors via the mixture approximation of Section~\ref{sec:33}, multi-criterion stopping rules and Bayesian posterior-probability decision rules. \pkg{gsbDesign} \citep{gerber2016,gerber2024}, which inspired the present work, provides closed-form normal--normal updates for continuous endpoints. \pkg{adaptr} \citep{granholm2022,granholm2026} is a general Bayesian adaptive-trial simulator with interim stopping, response-adaptive randomisation and multi-arm or platform support, but its conjugate-update implementation is limited to Gaussian and binomial outcomes. \pkg{BATSS} \citep{couturier2024,couturier2025} utilises \pkg{INLA} for GLM endpoints (continuous, binary and count), with broader parametric prior support than \pkg{gsbDesign} or \pkg{adaptr}, but does not provide a time-to-event interface and incurs substantially higher per-virtual-trial cost. It also supports response-adaptive randomisation and covariate adjustment, which are outside the current scope of \pkg{adabay}. Frequentist GSD packages like \pkg{rpact} \citep{wassmer2026} and \pkg{gsDesign} \citep{anderson2026} are mature and widely validated but do not target Bayesian posterior-probability decision rules and therefore cannot substitute for \pkg{adabay} when designs are calibrated to the operating characteristics induced by such rules.

\begin{table}[!ht]
\centering
\begin{tabular}{lccccccc}
\toprule
Package & Cont. & Bin. & Cnt. & TTE & Flexible prior & Multi-criteria & Bayes \\
\midrule
\pkg{adabay}     & \checkmark & \checkmark & \checkmark & \checkmark & \checkmark & \checkmark & \checkmark \\
\pkg{gsbDesign}  & \checkmark & --         & --         & --         & --         & --         & \checkmark \\
\pkg{adaptr}     & \checkmark & \checkmark & --         & --         & --         & --         & \checkmark \\
\pkg{BATSS}      & \checkmark & \checkmark & \checkmark & --         & --         & --         & \checkmark \\
\pkg{rpact}      & \checkmark & \checkmark & \checkmark & \checkmark & --         & --         & --         \\
\pkg{gsDesign}   & \checkmark & \checkmark & --         & \checkmark & --         & --         & --         \\
\bottomrule
\end{tabular}
\caption{Comparison of \pkg{adabay} with the closest existing tools. ``Cont.'' is continuous, ``Bin.'' is binary, ``Cnt.'' is count and ``TTE'' is time-to-event. ``Flexible prior'' refers to support for user-specified priors beyond the natural conjugate family, which \pkg{adabay} achieves through a finite-mixture-of-conjugate-components approximation with tail-probability diagnostics. The approximation covers any finite-mixture-approximable prior (Section~\ref{sec:33}), rather than every conceivable prior. \pkg{BATSS}'s \pkg{INLA} backend permits a choice among parametric prior families (Gaussian priors on the regression coefficients and \pkg{INLA}'s built-in hyperpriors) but not arbitrary user-specified priors, so its ``Flexible prior'' entry is ``--''. ``Multi-criteria'' indicates native support for stopping rules with two or more efficacy or futility criteria that must fire jointly at a look, for example, combining a non-superiority threshold and a clinical-relevance threshold. The \pkg{BATSS} TTE entry is ``--'' because \pkg{BATSS}'s sole model interface, \code{batss.glm}, is restricted to GLM likelihoods and cannot construct the censored event-time response (\code{inla.surv}) that an \pkg{INLA} time-to-event likelihood requires; this is a limitation of the \pkg{BATSS} interface, not of \pkg{INLA}, whose engine does provide exponential and other time-to-event likelihoods.}
\label{tab:761}
\end{table}

\section{Discussion}
\label{sec:8}
\pkg{adabay} provides an open-source \proglang{R} implementation of the semi-simulation framework of \citet{he2026} for fast evaluation and calibration of two-arm Bayesian GSDs across continuous, binary, count and time-to-event endpoints. It supports posterior-probability decision rules with one or more efficacy and futility criteria, binding or non-binding futility, flexible user-specified priors via finite conjugate-mixture approximations along with tail-probability diagnostics, and a precomputation strategy that decouples threshold calibration and look-time selection from the simulation pass. In the continuous and binary case studies, it reproduces the operating characteristics of \pkg{BATSS} and \pkg{adaptr} (and the analytic \pkg{gsbDesign} values in the continuous case) within Monte Carlo error, apart from the \pkg{adaptr} type~I error rate in the continuous case (Section~\ref{sec:71}). In the count case study, it reproduces the \pkg{BATSS} operating characteristics within Monte Carlo error under a recruited-patient person-time proxy for the exposure-driven look schedule, while running about four to five orders of magnitude faster than \pkg{BATSS} and one to over two orders of magnitude faster than \pkg{adaptr} per virtual trial at matched simulation budgets and reproducing the binary-endpoint benchmark of \citet{he2026}. For the time-to-event case study, no faithful comparator is constructible, since \pkg{BATSS}'s GLM-only \code{batss.glm} interface cannot construct a survival likelihood and \pkg{adaptr} does not cover that endpoint. This gap itself demonstrates that no single existing software spans all four endpoint types under the decision-rule and look-schedule conventions used here, whereas the semi-simulation framework itself is endpoint-agnostic, lowering the practical barrier to deploying Bayesian GSDs in confirmatory trials.

The semi-simulation framework, the per-endpoint posterior derivations and the convergence properties of the conjugate-mixture prior approximation are developed in detail in the companion methodology paper \citep{he2026}. The present paper focuses on the implementation and the programming interface of \pkg{adabay}, together with additional engineering features treated only abstractly in the methodology paper: the caching architecture, the parallel reproducibility guarantees, the tail-probability diagnostics for the prior approximation, and the four-endpoint unified interface. The two papers are therefore intended to be read together.

Three limitations should be noted. First, on the modelling side, the conjugate-mixture prior approximation is an additional source of error. Because the forward KL is mass-covering rather than tail-pinning, the tail-probability diagnostics of Section~\ref{sec:33} are necessary but not sufficient and should be inspected, with the operating characteristics re-checked at one more mixture component than the component-selection rule selects ($\hat{L}+1$; Section~\ref{sec:33}), before a calibrated design is relied on. The supported endpoint families are also tied to fixed outcome models, most notably the exponential survival model, which assumes a constant within-arm hazard. Second, on the computational side, calibration remains simulation-based and is therefore subject to Monte Carlo error. Regulatory-grade type I error precision \citep{fda2010,fda2026} requires at least $10^{5}$ trials per scenario, the computational cost scales unfavourably with the size of the design and threshold grids \citep{broglio2022}, and stochastic error makes principled optimisation of design parameters more difficult than in deterministic frequentist cases \citep{he2025}. The caching architecture (Section~\ref{sec:6}) and L'Ecuyer reproducibility (Section~\ref{sec:46}) mitigate but do not remove these issues. Third, the current design scope is deliberately restricted. Version~0.1.0 implements only two-arm fixed-allocation Bayesian GSDs across the four endpoint types in Table~\ref{tab:321}. Response-adaptive randomisation, covariate adjustment, multi-arm multi-stage and platform designs, arm-dropping, and sample-size re-estimation are not yet available. These are implementation choices rather than intrinsic limitations of the framework, and are addressed in the extensions below.

Several extensions are considered for future releases. On the computational side, beyond the caching architecture of Section~\ref{sec:6}, the most immediate gains come from common random numbers across data-generating values and across the threshold grid and from antithetic variates, which are essentially free and preserve the Monte Carlo error structure. Quasi-Monte Carlo sequences, single-instruction multiple-data vectorisation, reduced-precision arithmetic for tail probabilities far from the decision threshold, GPU offload of the trial loop, and boundary-targeted importance sampling would deliver larger reductions but either change the Monte Carlo error structure or undermine the strict single-threaded reproducibility guarantees of Section~\ref{sec:46}. A simulation-free or semi-analytical approach, analogous to the normal--normal route used by \pkg{gsbDesign}, would offer the largest reduction in computational cost. However, it does not generalise readily beyond natural conjugate settings without reintroducing substantial combinatorial cost over threshold calibration and look-time grids. 

On the methodological side, the roadmap for the next major release adds, in order, response-adaptive randomisation and sample-size re-estimation (both of which slot into the existing trial-simulation loop without altering the conjugacy-preserving update), covariate adjustment through a regression-augmented per-look posterior, multi-arm multi-stage and platform designs with optional arm-dropping, and a \pkg{Shiny}-based interactive interface for threshold calibration over the cached operating characteristics, along with a vignette-level prior-elicitation workflow building on the meta-analytic-predictive prior literature \citep{schmidli2014} to make informative priors easier to use at the design stage.

In summary, by bringing conjugate-speed posterior updating, flexible-prior support and a calibration-oriented caching architecture into a single open-source package, \pkg{adabay} lowers the computational barrier to using Bayesian group sequential designs at the planning stage of confirmatory trials.

\subsection*{Computational details}
The results in this work were obtained using \proglang{R}~4.5.0 with \pkg{adabay}~0.1.0, \pkg{adaptr}~1.5.0 \citep{granholm2026}, \pkg{BATSS}~1.1.1 \citep{couturier2025}, \pkg{INLA}~25.06.07 \citep{rue2009} and \pkg{RBesT}~1.8-2 \citep{weber2021} (whose underlying meta-analytic-predictive prior methodology is described in \citealt{schmidli2014}). All benchmarks were run on a single workstation with a 4.2~GHz quad-core Intel Core~i7 CPU (8 logical threads) and 32~GB of 2400~MHz DDR4 memory, under macOS, with each \pkg{adabay} scenario run on $8$ CPU cores via \code{mclapply} with the L'Ecuyer combined multiple-recursive generator (\code{cores = 8}), \pkg{adaptr} on $8$ cores via its native parallel backend, \pkg{BATSS} on $8$ \code{mclapply} workers with \pkg{INLA} pinned to one thread per worker, and \pkg{gsbDesign} on a single core.
\proglang{R} and all dependent packages on the Comprehensive R Archive Network are available at \url{https://CRAN.R-project.org/}. 
\pkg{adabay} is available at \url{https://github.com/trilixlab/adabay}, and the version of record at the time of writing is tagged \code{v0.1.0}.
\pkg{INLA} is available from the R-INLA project at \url{https://www.r-inla.org/}.

\section*{Acknowledgments}
We thank Suzie Cro and Laurent Billot for helpful discussions and constructive feedback.

\subsection*{Financial disclosure}
The authors declare no funding associated with the work presented in this article.

\subsection*{Conflict of interest}
The authors declare no potential conflicts of interest.

\bibliography{ZH2023_Manuscript}

\clearpage
\appendix

\section{Posterior of $\Delta$ for binary endpoints on relative-effect scales}
\label{apx:A}
For binary endpoints, the joint posterior of $(\vartheta_c,\vartheta_t)$ can factorise into two independent beta posteriors, $\vartheta_a\mid\mathcal{D}_k\sim\mathrm{Beta}(\tilde{a}_{a,k},\tilde{b}_{a,k})$, where $\tilde{a}_{a,k}=a_a+S_{a,k}$ and $\tilde{b}_{a,k}=b_a+n_{a,k}-S_{a,k}$ for $a\in\{c,t\}$, and $S_{a,k}$ is the cumulative number of responders in arm $a$ at look $k$. On the risk-ratio scale, with $\delta=\vartheta_t/\vartheta_c$, the change of variables from $(\vartheta_c,\vartheta_t)$ to $(\vartheta_c,\delta)$ has Jacobian $\vartheta_c$, and the posterior density of $\Delta$ is
\begin{linenomath}
\begin{equation}
\label{eqn:A1}
p_k(\delta\mid\mathcal{D}_k)
=
\int_{0}^{\min\{1,\,1/\delta\}}\vartheta\,
f(\vartheta;\,\tilde{a}_{c,k},\,\tilde{b}_{c,k})\,
f(\vartheta\delta;\,\tilde{a}_{t,k},\,\tilde{b}_{t,k})\,d\vartheta.
\end{equation}
\end{linenomath}
On the odds-ratio scale, with $\delta=[\vartheta_t/(1-\vartheta_t)]/[\vartheta_c/(1-\vartheta_c)]$, the transformation from $(\vartheta_c,\vartheta_t)$ to $(\vartheta_c,\delta)$ has Jacobian $\vartheta_c(1-\vartheta_c)/(1-\vartheta_c+\vartheta_c\delta)^2$, and the posterior density is
\begin{linenomath}
\begin{equation}
\label{eqn:A2}
p_k(\delta\mid\mathcal{D}_k)
=
\int_{0}^{1}\frac{\vartheta(1-\vartheta)}{(1-\vartheta+\vartheta\delta)^2}\,
f(\vartheta;\,\tilde{a}_{c,k},\,\tilde{b}_{c,k})\,
f\!\left(\frac{\vartheta\delta}{1-\vartheta+\vartheta\delta};\,\tilde{a}_{t,k},\,\tilde{b}_{t,k}\right)\,d\vartheta,
\end{equation}
\end{linenomath}
where $f(\vartheta;a,b)=\vartheta^{a-1}(1-\vartheta)^{b-1}/B(a,b)$ is the beta density with shape parameters $a,b$ and $B(a,b)$ is the beta function. Both integrals are one-dimensional and are evaluated by the same fixed-node Gauss--Legendre quadrature routine ($N=64$ nodes) that \pkg{adabay} uses for the risk-difference scale. In the implementation, \pkg{adabay} evaluates these tail probabilities in the algebraically equivalent form $\mathbb{P}(\vartheta_t>g(\vartheta_c)\mid\mathcal{D}_k)$, integrating the control-arm posterior density against the treatment-arm posterior tail with the appropriate boundary $g(\cdot)$ for each scale, which returns the identical value without forming the Jacobian explicitly.

\section{Posterior of $\Delta$ for count and time-to-event endpoints on relative-effect scales}
\label{apx:B}
For count and time-to-event endpoints, suppose the per-arm parameters have independent gamma posteriors, $\lambda_a\mid\mathcal{D}_k\sim\mathrm{Gamma}(\tilde{a}_{a,k},\tilde{b}_{a,k})$. The posterior distribution of the rate ratio (or hazard ratio) $\Delta=\lambda_t/\lambda_c$ admits a closed-form representation as a deterministic transformation of a beta random variable. With $U_a=\tilde{b}_{a,k}\lambda_a$ for $a\in\{c,t\}$, $U_a\sim\mathrm{Gamma}(\tilde{a}_{a,k},1)$, the standard ratio result for two independent gamma variables of unit rate gives $U_t/(U_c+U_t)\sim\mathrm{Beta}(\tilde{a}_{t,k},\tilde{a}_{c,k})$. Substituting $U_a=\tilde{b}_{a,k}\lambda_a$ and rearranging in $\Delta$ yields
\begin{linenomath}
\begin{equation}
\label{eqn:B1}
\frac{\Delta}{\Delta+\tilde{b}_{c,k}/\tilde{b}_{t,k}}
\sim
\mathrm{Beta}(\tilde{a}_{t,k},\,\tilde{a}_{c,k}).
\end{equation}
\end{linenomath}
The required tail probabilities are therefore obtained from the standard beta cumulative distribution function and require no numerical integration. The log rate-ratio (and log hazard-ratio) scales inherit the same closed form: the transformation is monotone, so $\mathbb{P}(\log\Delta>e\mid\,\cdot\,)=\mathbb{P}(\Delta>\exp(e)\mid\,\cdot\,)$ is again obtained from the beta distribution function, and no numerical integration is required.

\section{Expectation--maximisation updates for the conjugate-mixture prior}
\label{apx:C}
We summarise the expectation--maximisation updates that underlie the conjugate-mixture prior approximation. In the implementation, \code{fit\_mixture()} delegates the normal, beta and gamma kernels to \code{RBesT::automixfit()} when the suggested \pkg{RBesT} package is available, and otherwise applies the internal beta-kernel update below as a fallback; the normal--inverse-gamma kernel that \pkg{adabay} also supports is used for direct conjugate prior specification via \code{set\_prior()} rather than through \code{fit\_mixture()}. Throughout, $\psi^{(s)}$ denotes the $s$-th of $S$ samples drawn independently from the user-specified prior $p(\psi)$, where $\psi$ is the generic parameter introduced in Section~\ref{sec:33}. The latent indicator $z_{s,l}$ denotes membership of sample $s$ in component $l$. The E-step is common to all four kernels and computes the responsibilities
\begin{linenomath}
\begin{equation*}
\gamma_{s,l}^{(j)}
=
\frac{w_l^{(j)}\,h(\psi^{(s)};\,\boldsymbol{\eta}_l^{(j)})}{\sum_{l'=1}^{L}w_{l'}^{(j)}\,h(\psi^{(s)};\,\boldsymbol{\eta}_{l'}^{(j)})},
\end{equation*}
\end{linenomath}
where $w_l^{(j)}$ and $\boldsymbol{\eta}_l^{(j)}$ denote the weight and natural parameters of component $l$ at iteration $j$. The M-step updates $\boldsymbol{\eta}_l$ by maximum likelihood within each kernel, in closed form for the normal kernel, in which $\boldsymbol{\eta}_l=(\nu_l,\rho_l^2)$ is updated to the responsibility-weighted mean and variance, and numerically for the beta and gamma kernels, whose score equations are not closed form. When \pkg{RBesT} is unavailable, the internal beta-kernel fallback replaces the maximum-likelihood M-step by a responsibility-weighted method-of-moments update, and declares convergence when the maximum absolute change in the mixture weights and component parameters falls below $10^{-6}$ between consecutive iterations, with a hard cap of $200$ iterations; the normal and gamma kernels are unavailable on that path. The implementation follows \citet{dempster1977} and the standard textbook treatment in \citet{mclachlan2008}. We omit the routine algebra here and refer the reader to the package source.

\end{document}